\documentclass[10pt]{article} 
\usepackage[utf8]{inputenc} 
  \usepackage{graphicx}

\usepackage{geometry} 
\usepackage{graphicx} 
\usepackage{booktabs} 
\usepackage{array} 
\usepackage{paralist} 
\usepackage{verbatim} 
\usepackage{subfig} 
\usepackage{amsmath,epsfig}
\usepackage[greek,english]{babel}
\usepackage{multicol}
\usepackage[font={small,it}]{caption}
\usepackage{physics}
\usepackage{gensymb}
\usepackage{hyperref}
\usepackage{dutchcal}
\usepackage{authblk}
\usepackage{sectsty}

\usepackage{fancyhdr} 
\usepackage{sectsty}
\allsectionsfont{\sffamily\mdseries\upshape} 
\usepackage[nottoc,notlof,notlot]{tocbibind} 
\usepackage[titles,subfigure]{tocloft} 

\title{Carter's case [$\mathcal{D}$] admits the 2nd Canonical Form of the Killing Tensor}
\author[1]{D. Kokkinos}
\author[2]{T. Papakostas}
\affil[1]{Department of Information and Communication System Engineering, University of Aegean, Karlovassi, Samos, Greece}
\affil[2]{Department of Electrical and Computer Engineering, Hellenic Mediterranean University, Heraklion, Crete, Greece }
\begin{document}
\maketitle

\begin{abstract}
The study of the Canonical Forms of the Killing Tensor concerns the simultaneous resolving of the Integrability Conditions of the Killing Tensor along with the Einstein's Field Equations employing the framework of Newman-Penrose Formalism. We present all the Petrov Types admitting the 2nd and 3rd Canonical Forms of Killing Tensor in Vacuum in the frame of General Theory of Relativity. During the investigation of the Type D solution of 2nd Canonical form of the Killing Tensor the Carter's Case [$\mathcal{D}$] solution in Vacuum emerged.   
\end{abstract}



\section{Introduction}	

In order to find exact solutions of Einstein's Field Equations additional hypothesis and mathematical assumptions must be used. Symmetries have a crucial role during the resolving procedure of the equations and could be implied with various ways. We choose to deploy the Killing Tensor since its existence in the context of General Theory of Relativity lies with the physical concept of the problem. The admission of the Penrose-Floyd tensor is correlated with the conservation of Laplace-Runge-Lenz vector of the Kepler problem in geodesics as a constant of motion \cite{cariglia2014hidden} ,  \cite{taxiarchis1985space} since the relation $K_{\mu \nu} p^\mu p^\nu$ is a constant of motion if $p^\mu$ is the momentum of an observer during the geodesic motion\footnote{The indices $\mu,\nu$ take the values 1,2,3,4.}.

Additionally the study of the Killing tensor in the context of General Theory of Relativity could be proved fruitful, providing spacetimes with hidden symmetries. These kind of symmetries emerge during the study of the dynamics of a system featuring the conserved quantities of the system or one-parameter isometries which is equivalent with the admission of existence of Killing vectors.

 Any tensor of order 2 that satisfies the Killing Equation is called a Killing tensor.
 $$K_{(\mu \nu ; \alpha)} =0$$
 Hence, the function -in a broad sense- of the Integrability Conditions of the Killing Tensor are the auxilliary relations that one could use to confront these kind of problems. Actually the usage of the integrability conditions of a Killing tensor could be metaphorically described as a double-edged sword. In one hand we have to deal with an overdetermined system of equations and finally to obtain a solution but on the other hand maybe this could be practically paves the way with quite special solutions.

Anyway the main idea that motivated us to deal with the Canonical Forms of the Killing tensor in the frame of General Relativity was \textit{that we might find new interesting spacetimes in vacuum if we deal with more general forms of a Killing tensor with more than two distinct eigenvalues}. During the last decades only few vacuum solutions have obtained which admit a Killing Tensor. This work done by Hauser-Malhiot \cite{hauser1976space} for the Killing tensor of Segré type [(11),(1,1)] and the result was one of the most general family of stationary axially symmetric electro-vacuum spacetimes that found independently by Carter \cite{carter1968hamilton}. Regarding the topic just mentioned the Killing Tensor that has been studied by Hauser-Malhiot is a special case of our Canonical Forms.
 
There are four Canonical Forms of Killing tensor in the context of GR which admit one timelike and three spacelike eigenvectors. The following Canonical forms of the Killing Tensor are presented as covariant tensors. In order to obtain the eigenvalues and the eigenvectors of each form, mixed tensor must be obtained using the metric tensor. The $\lambda$'s are the decomposed eigenvalues since are the elements which embody the eigenvalues. Consider the latter we are going to refer to $\lambda$'s as eigenvalues for reasons of simplification. 

 \textbf{The first form $K^1_{\mu \nu}$ is a Jordan Canonical form with one double eigenvalue $\lambda_1$ (Jordan Block) and two others which are real $-(\lambda_2 \pm \lambda_7)$, the second canonical form $K^2_{\mu \nu}$ have 2 real eigenvalues $\lambda_0\pm \lambda_1$ and the same other two real eigenvalues $-(\lambda_2 \pm \lambda_7)$, while $K^3_{\mu \nu}$ have two complex conjugate $\lambda_0 \pm i \lambda_1  $ and other two real eigenvalues which are the same with the previous forms. Finally the $K^4_{\mu \nu}$ form has one double eigenvalue $\lambda_1$ (Jordan Block) and the other two real eigenvalues are the same with all the previous forms}.

$$ {K{^1}}_{\mu \nu} =  \begin{pmatrix}
0 & \lambda_1   & 0 & 0\\
 \lambda_1 & \lambda_0 & 0 & 0\\
0 & 0&  \lambda_7   &\lambda_2 \\
0 & 0 & \lambda_2 & \lambda_7   
\end{pmatrix} {K{^2}}_{\mu \nu} = \begin{pmatrix}
\lambda_0 & \lambda_1   & 0 & 0\\
 \lambda_1 & \lambda_0 & 0 & 0\\
0 & 0&  \lambda_7   &\lambda_2 \\
0 & 0 & \lambda_2 & \lambda_7   
\end{pmatrix} {K{^3}}_{\mu \nu} = \begin{pmatrix}
\lambda_0 & \lambda_1   & 0 & 0\\
 \lambda_1 & -\lambda_0 & 0 & 0\\
0 & 0&  \lambda_7   &\lambda_2 \\
0 & 0 & \lambda_2 & \lambda_7   
\end{pmatrix} $$

$$ {K{^4}}_{\mu \nu} = \begin{pmatrix}
0 & \lambda_1   & -p & -\bar{p} \\
 \lambda_1 & 0 & 0 & 0\\
-p & 0&  \lambda_7   &\lambda_2 \\
-\bar{p} & 0 & \lambda_2 & \lambda_7   
\end{pmatrix} \hspace{1cm} p, \bar{p} = \pm 1$$

It should be noted that the only difference between $K^2$ and $K^3$ could be described via a factor $q$. We choose to deal simultaneously with forms $K^2$ and $K^3$ with the usage of the parameter $q=\pm 1$ that gives us the 2nd and 3rd forms for $+1$ and $-1$ accordingly. Also we choose to annihilate the eigenvalue $\lambda_7$  for reasons that will be revealed later.  

The setup of the problem involves the treatment of the system of equations that must be satisfied along with the necessary restrictions of the problem. The restrictions we choose pave the way for algebraically special solutions according to Petrov classification \cite{petrov2000classification}. Also we demand one eigenvalue at most to be constant and neither of them zero as well as we don't scope to deal with solutions of Type O Solution or Conformally Flat Spacetimes (CF). 

\section{Methodology}

First things first, in the \textbf{Section 3} we describe the main points of the appointed formalism. The Newman-Penrose Formalism it is a widely known formalism which use null tetrads and was presented by Newman and Penrose \cite{newman1962approach} and analyzed geometrically by Debever \cite{debeverriemann} and Cahen and et. al \cite{cahen1967complex}. It was found in order to describe the gravitational radiation in the frame of GR but proved to have much more usefulness. 

We initiate the treatment of the problem in \textbf{Section 4}, where we are going to unravel the setup of the problem. The main steps of the resolving procedure concerns the collection of the Killing equations and the production of the Integrability Conditions, which we able to obtain via the commutation relations of the directional derivatives of the formalism. The Killing equations dictate the initial spins annihilation and expressions that correlate the spin themselves. In the same time we demand by the eigenvalues of the Killing tensor to be general, so any unnecessary annihilation of the directional derivatives of the eigenvalues is not a desired outcome. 

 Next a rotation around our null base tetrads is implied since it seems unbearable for someone to extract any results without further simplifications. The latter provides us with the \textbf{Key relations}. Based upon the latter we resulting to the geometric classification of the emerged solutions, the Petrov Classification denotes the geometric nature of the solution we have to solve. 
 
Till this point our investigation based in the simultaneous admitting of both Killing Tensors $K^2,K^3$. In the \textbf{section 5} though we are going to give up with this simultaneous admission of both Killing tensors and we will consider a relation between the spin coefficients. The lattter forbids the existence of $K^3$ tensor (q=-1). With the implication of Frobenius Theorem of integrability we set up our vector basis which depend by random functions. The determination of these functions finally resulting to Carter's case [$\mathcal{D}$]. 

The next section referred to the analysis of the solution which takes place in \textbf{Section 6}. Scoping to gain physical knowledge about the solution necessary reductions have to be operated on . At the last section the equations of geodesics are presented along with the characterization of the eigenvalues of the Killing Tensor in respect to constants of motion. Finally the \textbf{Appendices A,B} demonstrate proofs that would impede the flow of our syllogism if we emplaced them in the main body of the article.   

\section{Newman-Penrose Formalism}

General Relativity is based on the concept of spacetime which is 4-dimensional differentiable manifold endowed with a Lorentzian metric $g_{\mu \nu}$.  The four vectors  form a covariant null tetrad and the metric metric can be put in the form 

$$ds^2 = 2(\theta^1 \theta^2 - \theta^3 \theta^4) $$

\vspace{0.5cm}

where $\theta$'s are four independent Pfaffian forms in their general forms defined by the following relations. These four vectors form a covariant null orthogonal tetrad (Greek indices = 1,2,3,4, Latin indices =1,2,3).

$$ \theta^1 = n_\mu dx^\mu \hspace{0.8cm} \theta^2 = l_\mu dx^\mu \hspace{0.8cm} \theta^3 = - \bar{m}_\mu dx^\mu \hspace{0.8cm} \theta^4 = - m_\mu dx^\mu  $$

\vspace{0.5cm}

Where the general metric $g_{\mu \nu}$ is 

$$g_{\mu \nu} = l_\mu n_\nu + n_\mu l_\nu              - m_\mu \bar{m}_\nu - \bar{m}_\mu m_\nu  $$

\vspace{0.5cm}

The orthogonality properties of the vectors are

$$l_\mu l^\mu = m_\mu m^\mu = \overline{m}_\mu \bar{m}^\mu = n_\mu n^\mu = 0 $$
$$l_\mu n^\mu = 1 = - m_\mu \bar{m}^\mu $$
$$ l_\mu m^\mu = l_\mu \bar{m}^\mu = n_\mu m^\mu = n_\mu \bar{m}^\mu = 0$$

\vspace{0.5cm}

The directional derivative that operated on the field equations of the formalism are given by

$$D \phi  \equiv \phi_{, \mu } n^{\mu} $$
$$\Delta \phi \equiv \phi_{, \mu } l^{\mu} $$
$$\delta \phi  \equiv \phi_{, \mu } m^{\mu} $$
$$ \bar{ \delta}\phi \equiv \phi_{, \mu } \bar{ m }^{\mu} $$

\vspace{0.5cm}

A basis for the space ($C_3$) of complex self-dual bivector (2-forms) is given by

$$Z^1 = \theta^1 \wedge \theta^3 \hspace{0.8cm} Z^2 = \theta^1 \wedge \theta^2 - \theta^3 \wedge \theta^4 \hspace{0.8cm} Z^3 =  \theta^4 \wedge \theta^2  $$

\vspace{0.5cm}
The composite of the metric in this base are 

$$ \gamma^{a b} = 4({\delta^{a}}_{(1} {\delta^b}_{3)} -  {\delta^{a}}_{(2} {\delta^b}_{2)} ) $$

\vspace{0.5cm}

The complex connection 1-forms $\sigma_{b}^a$ are defined by 

$$dZ^a = \sigma^{a}_b \wedge Z^b $$

\vspace{0.5cm}

The vectorial connection 1-form is defined by 

$$\sigma_a = \frac{1}{8} e_{a b c} \gamma^{c d} \sigma^{b}_{d} $$

Where $ e_{a b c} $ is the Levi-Civita tensor. The tetrad components $ \sigma_{a \alpha} $ defined by $ \sigma_{a} = \sigma_{a \alpha}  \theta^{\alpha}$ that are 12 complex valued functions which are the 12 complex spin coefficients.

$$\sigma_{a \alpha} = \begin{bmatrix}  \kappa & \tau & \sigma & \rho \\ \epsilon & \gamma & \beta & \alpha \\ \pi & \nu & \mu & \lambda\\          \end{bmatrix}$$

The following forms denote the relations between spins with tetrads.

$$\rho = l_{\mu;\nu} m^\mu \bar{m}^\nu$$
$$\lambda =  -n_{\mu;\nu} \bar{m}^\mu \bar{m}^\nu $$
$$\kappa = l_{\mu ; \nu} m^\mu l^\nu$$
$$\pi = -n_{\mu;\nu} \bar{m}^\mu l^\nu$$
$$\beta = \frac{1}{2}(l_{\mu ; \nu} n^\mu m^\nu - m_{\mu ; \nu} \bar{m}^\mu m^\nu )$$
$$\epsilon = \frac{1}{2} (l_{\mu ; \nu} n^\mu l^\nu - m_{\mu ; \nu} \bar{m}^\mu l^ \nu) $$

 $$ \sigma = l_{\mu ; \nu} m^\mu m^\nu$$
  $$\mu = -n_{\mu;\nu} \bar{m}^\mu m^\nu $$
$$\tau =   l_{\mu ; \nu} m^\mu n^\nu$$
$$ \nu =  -n_{\mu;\nu} \bar{m}^\mu n^\nu   $$
$$ \alpha = \frac{1}{2}(l_{\mu ; \nu} n^\mu \bar{m}^\nu - m_{\mu ; 
\nu} \bar{m}^\mu \bar{m}^\nu )$$
$$\gamma = \frac{1}{2} ( l_{\mu ; \nu} n^\mu n^\nu - m_{\mu ; \nu}\bar{m}^\mu n^\nu) $$

\vspace{0.5cm}

The complex curvature 2-forms $ \Sigma^{b}_{d} $ are defined by 

$$ \Sigma^{b}_{d} = d \sigma_{d}^{b} + \sigma^{b}_{g} \wedge \sigma^{g}_{d}   $$

\vspace{0.5cm}

and the vectorial curvature 2-form by 

$$ \Sigma_{a} = \frac{1}{8} e_{a b g } \gamma^{g d } \Sigma{^b}_d $$

\vspace{0.5cm}

On expanding $\Sigma_a$ in the basis of $ \left[ Z^a , \bar{Z}^a \right]$ we can obtain

$$\Sigma_a = (C_{a b} - \frac{1}{6} R \gamma_{ a b})Z^b + E_{a b} \bar{Z}^b  $$ 

\vspace{0.5cm}

where these quantities are related with the curvature components of the formalism. 

$$ C_{ a b} = \begin{bmatrix}  \Psi_0 & \Psi_1 & \Psi_2 \\ \Psi_1 & \Psi_2 &\Psi_3 \\ \Psi_2 & \Psi_3 & \Psi_4\\          \end{bmatrix},  \hspace{1cm} E_{ a b} = \begin{bmatrix}  \Phi_{00} & \Phi_{01} & \Phi_{02} \\ \Phi_{10} & \Phi_{11} &\Phi_{12} \\ \Phi_{20} & \Phi_{21} & \Phi_{22}\\          \end{bmatrix}$$

\vspace{1cm}

In this formalism the 10 Weyl's components are represented by the 5 complex scalar functions

$$ \Psi_0 = C_{\kappa \lambda \mu \nu} k^\kappa m^\lambda k^\mu m^\nu $$
$$ \Psi_1 = C_{\kappa \lambda \mu \nu} k^\kappa l^\lambda k^\mu m^\nu $$
$$ \Psi_2 = C_{\kappa \lambda \mu \nu} k^\kappa m^\lambda \bar{m}^\mu l^\nu $$
$$ \Psi_3 = C_{\kappa \lambda \mu \nu} l^\kappa k^\lambda l^\mu \bar{m}^\nu $$
$$ \Psi_4 = C_{\kappa \lambda \mu \nu} l^\kappa \bar{m}^\lambda l^\mu \bar{m}^\nu $$
\vspace{1cm}

The Ricci tensor components represented by $E_{a b}$ and they are divided to real and imagine components. The real components are those with the same index.

\vspace{0.5cm}

$$\Phi_{00} = \frac{1}{2} R_{\mu \nu} k^\mu k^\nu$$
$$\Phi_{11} = \frac{1}{4} R_{\mu \nu} (k^\mu l^\nu + m^\mu \bar{m}^\nu)$$
$$\Phi_{22} = \frac{1}{2} R_{\mu \nu} l^\mu l^\nu $$

\vspace{0.5cm}

The complex components are given by the following relations and constrained by $\Phi_{AB} = \bar{\Phi}_{BA}$.

\vspace{0.3cm}

$$\Phi_{01} = \frac{1}{2} R_{\mu \nu} k^\mu m^\nu$$
$$\Phi_{02} = \frac{1}{2} R_{\mu \nu} m^\mu m^\nu$$
$$\Phi_{12} = \frac{1}{2} R_{\mu \nu} l^\mu m^\nu$$

\vspace{1cm}

All these quantities describe the main parts of the Einstein's Field Equations. The EFE in this formalism are represented by the correspond field equations which are known either as Newman-Penrose Field Equations or as Ricci identities \cite{newman1962approach}. These equations relate the spin-coefficients to the derivatives of the tetrad components, the spin-coefficient equations describe the relationship of the curvature tensor to derivatives of the connection (the spin-coefficients). A novelty here is that these equations are integrated not one set at a time, but together, i.e., by going back and forth between the sets. The following relations are presented without the spin coefficients $\sigma$ and $\lambda$ since in our case are annihilated from the start.

\vspace{0.5cm}
The Newman-Penrose Field Equations

\begin{equation}\tag{a} D \rho  - \bar{\delta} \kappa = {\rho}^2 + \rho ( \epsilon + \bar{\epsilon}) - \bar{\kappa} \tau - \kappa (2(\alpha +\bar{\beta}) + (\alpha - \bar{\beta}) - \pi) \end{equation}
\begin{equation}\tag{b} \delta \kappa = \kappa ( \tau - \bar{\pi} +2(\bar{\alpha} +\beta) - (\bar{\alpha} - \beta) ) - \Psi _o  \end{equation}
\begin{equation}\tag{c} D\tau = \Delta \kappa + \rho(\tau + \bar{\pi})+ \tau(\epsilon - \bar{\epsilon})    -2\kappa \gamma - \kappa (\gamma + \bar{\gamma}) + \Psi_1 \end{equation}
\begin{equation}\tag{i} D\nu - \Delta \pi = \mu(\pi + \bar{\tau}) +\pi(\gamma - \bar{\gamma}) -2\nu \epsilon - \nu (\epsilon + \bar{\epsilon}) +\Psi_3 \end{equation}
\begin{equation}\tag{g} \bar{\delta} \pi = - \pi(\pi + \alpha - \bar{\beta}) + \nu\bar{\kappa} \end{equation}
\begin{equation}\tag{p} \delta \tau = \tau (\tau - \bar{\alpha} + \beta) -\bar{\nu} \kappa  \end{equation}
\begin{equation}\tag{h} D\mu - \delta\pi = \mu \bar{\rho} + \pi(\bar{\pi} - \bar{\alpha} +\beta) -\mu (\epsilon +\bar{\epsilon}) - \kappa \nu + \Psi_2 + 2\Lambda\end{equation}
\begin{equation}\tag{n} \delta\nu - \Delta \mu = \mu (\mu + \gamma + \bar{\gamma}) - \bar{\nu} \pi + \nu (\tau - 2(\bar{\alpha}+\beta) +(\bar{\alpha} -\beta) )\end{equation}
\begin{equation}\tag{q} \Delta \rho - \bar{\delta} \tau = - \bar{\mu} \rho - \tau(\bar{\tau} + \alpha - \bar{\beta}) + \nu \kappa + \rho(\gamma + \bar{\gamma}) - \Psi_2 - 2\Lambda\end{equation}
\begin{equation}\tag{k} \delta\rho = \rho(\bar{\alpha} + \beta) +\tau(\rho-\bar{\rho})+ \kappa(\mu-\bar{\mu}) - \Psi_1 \end{equation}
\begin{equation}\tag{m} \bar{\delta} \mu = -\mu (\alpha +\bar{\beta}) -\pi (\mu- \bar{\mu}) - \nu (\rho-\bar{\rho}) + \Psi_3 \end{equation}
\begin{equation}\tag{d} D\alpha - \bar{\delta} \epsilon = \alpha(\rho + \bar{\epsilon} -2\epsilon) - \bar{\beta}\epsilon - \bar{\kappa}\gamma + \pi (\epsilon + \rho)\end{equation} 
\begin{equation}\tag{e} D \beta - \delta{\epsilon} = \beta(\bar{\rho} - \bar{\epsilon}) -\kappa(\mu + \gamma) -\epsilon(\bar{\alpha} - \bar{\pi}) + \Psi_1\end{equation} 
\begin{equation}\tag{r} \Delta \alpha - \bar{\delta}\gamma = \nu(\epsilon+\rho) +\alpha( \bar{\gamma} - \bar{\mu}) +\gamma (\bar{\beta}- \bar{\tau}) - \Psi_3 \end{equation}
\begin{equation}\tag{o} -\Delta \beta + \delta \gamma = \gamma(\tau -\bar{\alpha} - \beta) +\mu \tau - \epsilon \bar{\nu} - \beta( \gamma - \bar{\gamma} -\mu)\end{equation}
\begin{equation}\tag{l} \delta \alpha - \bar{\delta}\beta = \mu \rho +\alpha (\bar{\alpha} - \beta) - \beta(\alpha - \bar{\beta})+  \gamma(\rho - \bar{\rho}) +\epsilon (\mu-\bar{\mu})-\Psi_2 + \Lambda\end{equation}
\begin{equation}\tag{f} D\gamma - \Delta \epsilon = \alpha(\tau + \bar{\pi}) + \beta( \bar{\tau} + \pi) - \gamma ( \epsilon+\bar{\epsilon}) - \epsilon (\gamma +\bar{\gamma})\end{equation}
\begin{equation}\tag{j} \bar{\delta}\nu = - \nu(2(\alpha +\bar{\beta} ) + (\alpha - \bar{\beta})  + \pi - \bar{\tau}  ) + \Psi_4\end{equation}

\vspace{0.5cm}
The Bianchi Identities

\begin{equation}\tag{I} \bar{\delta} \Psi_0 - D \Psi_1 = (4\alpha - \pi )\Psi_0  - 2(2\rho +\epsilon)\Psi_1 +3\kappa \Psi_2\end{equation}
\begin{equation}\tag{II} \bar{\delta} \Psi_1 -D\Psi_2 = 2(\alpha - \pi)\Psi_1 -3\rho \Psi_2 +2\kappa \Psi_3 \end{equation}
\begin{equation}\tag{III} \bar{\delta}\Psi_2 - D\Psi_3 = -3\pi \Psi_2 +2(\epsilon - \rho) \Psi_3 +\kappa\Psi_4\end{equation}
\begin{equation}\tag{IV} \bar{\delta}\Psi_3 - D\Psi_4 = -2(\alpha +2\pi)\Psi_3 +(4\epsilon - \rho)\Psi_4 \end{equation} 
\begin{equation}\tag{V} \Delta \Psi_0 - \delta \Psi_1 = (4\gamma - \mu) \Psi_0 -2(2\tau +\beta)\Psi_1\end{equation}
\begin{equation}\tag{VI} \Delta \Psi_1 - \delta \Psi_2 = \nu \Psi_0 + 2(\gamma - \mu)\Psi_1  -3\tau\Psi_2\end{equation}
\begin{equation}\tag{VII} \Delta \Psi_2 - \delta \Psi_3 = 2\nu\Psi_1 -3\mu\Psi_2 +2(\beta-\tau)\Psi_3\end{equation} 
\begin{equation}\tag{VIII} \Delta \Psi_3 - \delta \Psi_4 = 3\nu \Psi_2 - 2(\gamma+2\mu) \Psi_3 +(4\beta-\tau)\Psi_4\end{equation} 

Also the Lie bracket plays an important role to the theory in order to obtain the commutation relations of the NP formalism. The CR come to surface with the usage of the Lie bracket with the vectors $n^\mu, l^\mu, m^\mu,\bar{m}^\mu $. For example 
\begin{equation}\tag{CR1} [n^\mu,l^\mu ] = [D,\Delta] =  (\gamma+\bar{\gamma})D + (\epsilon + \bar{\epsilon})\Delta - (\pi + \bar{\tau})\delta - (\bar{\pi} +\tau)\bar{\delta} \end{equation} 
Hence, we have four commutations relations (CR) in every possible combination of the vector basis. We present it here with every detail divided into real and imaginary parts implying that $\sigma = \lambda = 0$.
\begin{equation}\tag{$CR2_+$} [(\delta+\bar{\delta}),D] = (\alpha + \bar{\alpha} + \beta + \bar{\beta} - \pi - \bar{\pi} )D + (\kappa+\bar{\kappa})\Delta - (\bar{\rho}+\epsilon -\bar{\epsilon})\delta - (\rho - \epsilon + \bar{\epsilon})\bar{\delta}  \end{equation}
\begin{equation}\tag{$CR2_-$} [(\delta-\bar{\delta}),D] = (-\alpha + \bar{\alpha} + \beta - \bar{\beta} + \pi - \bar{\pi} )D + (\kappa-\bar{\kappa})\Delta - (\bar{\rho}+\epsilon -\bar{\epsilon})\delta + (\rho - \epsilon + \bar{\epsilon})\bar{\delta}  \end{equation}

\begin{equation}\tag{$CR3_+$}[(\delta+\bar{\delta}),\Delta] = -(\nu +\bar{\nu})D + (\tau+\bar{\tau} - \alpha - \bar{\alpha} - \beta - \bar{\beta})\Delta +(\mu -\gamma +\bar{\gamma})\delta +(\bar{\mu} +\gamma - \bar{\gamma})\bar{\delta} \end{equation}
\begin{equation}\tag{$CR3_-$}[(\delta-\bar{\delta}),\Delta] = -(\nu -\bar{\nu})D + (\tau-\bar{\tau} + \alpha - \bar{\alpha} - \beta + \bar{\beta})\Delta +(\mu -\gamma +\bar{\gamma})\delta -(\bar{\mu} +\gamma - \bar{\gamma})\bar{\delta} \end{equation}

\begin{equation}\tag{$CR4$} [\delta,\bar{\delta}] = -(\mu - \bar{\mu})D - (\rho-\bar{\rho}) \Delta + (\alpha - \bar{\beta})\delta - (\bar{\alpha} - \beta)\bar{\delta}\end{equation}

These are the Newman Penrose Field Equations, the Bianchi Identities and the commutation relations of the basis vectors with $\sigma = \lambda = 0$. NPEs is a set of 18 linear equations in comparison with the non-linear 10 equations of 3-1 formalism. Despite the fact that we have to solve a considerably larger number of equations in comparison with 3-1 formalism, when we use coordinates directly this formalism has great advantages. All differential equations are of first order. Also gauge transformations of the tetrad can be used to simplify the field equations. One can easily extract invariant properties of the gravitational field without using a coordinate basis \cite{stephani2009exact}. 
 
The usage of this formalism allows the concentration of the equations on individual 'scalar' equations with particular physical or geometric significance, one thus could see a natural hierarchical structure in the set of Einstein equations, also it allows us to search for solutions with specific special features, such as the presence of one or two null directions that might be singled out by physical or geometric considerations. In conclusion, Newman-Penrose formalism, at the first sight, looks like a complex tool but it is proved to be sophisticated and convenient at last.

\section{Problem Setup}
Generally speaking in order to find an exact solution of NP the Newman-Penrose Equations (NPE) and the Bianchi Identities (BI) must be solved along with additional relations which are imposed by either geometrical characteristics or by symmetries such in our case. The Commutation Relations (CR) are very useful in an effort to extract expressions that connects the spin coefficients themselves. In our case the additional conditions are the Integrability Conditions (IC) of the eigenvalues $\lambda_0$, $\lambda_1$, $\lambda_2$ of our Killing tensor which come by the Killing Equation. 

It is worth noting that the Integrability Conditions frame the problem and, by extension, add the necessary `information' determining the solution. Along these lines we begin with the Killing equation. \begin{equation} K_{(\mu \nu ; \alpha)}= 0 \end{equation}

As one could see it is helpful to define the factor $q= \pm 1 $ because the only difference in the $K^2$ and $K^3$ forms is the $-1$ in the $K^3_{22}$ element of the tensor. We get the $K^2_{\mu \nu}$  for the $q=+1$ and the $K^3_{\mu \nu}$ for the $q=-1$. This modification allows us to solve the problem and in the same time gain a double solution which admits two Killing tensors. 

\begin{equation} K_{(\mu \nu ; \alpha)}= 0 \end{equation}

\begin{equation}K^{2,3}_{\mu \nu} = \lambda_0[n_\mu n_\nu + q l_\mu l_\nu]+ \lambda_1[l_\mu n_\nu + n_\mu l_\nu ]     - \lambda_2 [ m_\mu \bar{m}_\nu +\bar{m}_\mu m_\nu]\end{equation}

The covariant derivatives of every eigenvalue and of the vectors are the following.

\begin{equation}d \lambda = D\lambda \theta^1 +\Delta \lambda \theta^2 +\delta \lambda \theta^3 +\bar{\delta} \lambda \theta^4 \end{equation}

\begin{multline}\nabla_\alpha n_\mu = -(\epsilon + \bar{\epsilon})n_\alpha n_\mu -(\gamma+\bar{\gamma})l_\alpha n_\mu +(\alpha+\bar{\beta})m_\alpha n_\mu + (\bar{\alpha}+\beta)\bar{m}_\alpha n_\mu + \pi n_\alpha m_\mu\\
 + \nu l_\alpha m_\mu -\lambda m_\alpha m\mu  -\mu \bar{m}_\alpha m_\mu  +\bar{\pi}  n_\alpha \bar{m}_\mu +\bar{\nu} l_\alpha \bar{m}_\mu -\bar{\mu} m_\alpha \bar{m}_\mu -\bar{\lambda} \bar{m}_\mu \bar{m}\nu\end{multline}

\begin{multline}\nabla_\alpha l_\mu = (\epsilon + \bar{\epsilon})n_\alpha l_\mu +(\gamma+\bar{\gamma})l_\alpha l_\mu -(\alpha+\bar{\beta})m_\alpha l_\mu - (\bar{\alpha}+\beta)\bar{m}_\alpha l_\mu - \bar{\kappa}n_\alpha m_\mu \\
- \bar{\tau}l_\alpha m_\mu + \bar{\sigma} m_\alpha m\mu +\bar{\rho} \bar{m}_\alpha m_\mu  - \kappa n_\alpha \bar{m}_\mu -\tau l_\alpha \bar{m}_\mu +\rho m_\alpha \bar{m}_\mu + \sigma \bar{m}_\mu \bar{m}\nu\end{multline}

\begin{multline}\nabla_\alpha m_\mu = - \kappa n_\alpha n_\mu - \tau l_\alpha n_\mu +\rho m_\alpha n_\mu +\sigma \bar{m}_\alpha n_\mu +\bar{\pi} n_\alpha l_\mu +\bar{\nu} l_\alpha l_\mu - \bar{\mu} m_\alpha l_\mu \\
- \bar{\lambda} \bar{m}_\alpha l_\mu +(\epsilon -\bar{\epsilon})n_\alpha m_\mu +(\gamma - \bar{\gamma})l_\alpha m_\mu - (\alpha - \bar{\beta})m_\alpha m_\mu +(\bar{\alpha} - \beta)\bar{m}_\alpha m_\mu
\end{multline}

The Killing equation provide us with the following relations which will be used along with the CR in order to give us the IC of the eigenvalues. 

\begin{equation}\sigma = \lambda = 0\end{equation}

The directional derivatives for eigenvalue $\lambda_0$, $\lambda_1$, $\lambda_2$ proved to be the following

\begin{equation}D \lambda_0 = 2 \lambda_0 (\epsilon + \bar{\epsilon}) \end{equation}
\begin{equation}\Delta \lambda_0 =   -2\lambda_0 (\gamma + \bar{\gamma}) \end{equation}  
\begin{equation}\delta \lambda_0 =  2(\lambda_0 (\bar{\alpha} + \beta + \bar{\pi} ) - \kappa (\lambda_1 + \lambda_2) ) \end{equation}
\begin{equation}\delta \lambda_0 =  2( -\lambda_0 (\bar{\alpha} + \beta + \tau ) +q\bar{\nu} (\lambda_1 + \lambda_2) ) \end{equation}
\begin{equation}\delta \lambda_0 =  \lambda_0 (\bar{\pi} -\tau) -(\kappa -q\bar{\nu} )(\lambda_1 +\lambda_2)  \end{equation}

\begin{equation}D\lambda_1 = 2 \lambda_0 (\gamma + \bar{\gamma}) \end{equation}
\begin{equation}\Delta \lambda_1 = -2 q \lambda_0 (\epsilon + \bar{\epsilon}) \end{equation}
\begin{equation}\delta \lambda_1 =-q\lambda_0 (\kappa - q \bar{\nu}) +(\lambda_1 +\lambda_2)(\bar{\pi} - \tau)\end{equation}

\begin{equation}D\lambda_2 = \lambda_0 (\mu + \bar{\mu}) -(\lambda_1 +\lambda_2) (\rho +\bar{\rho}) \end{equation}
\begin{equation} \Delta \lambda_2 = -q\lambda_0 (\rho +\bar{\rho})   -(\lambda_1 +\lambda_2)   (\mu + \bar{\mu})\end{equation}
\begin{equation}  \delta \lambda_2 = 0\end{equation}

From (11), (12), (13) it is obvious that we can define the factor Q. Also if we add (11), (12) we take (13).

\begin{equation}Q \equiv \frac{\lambda_0}{\lambda_1 +\lambda_2} = \frac {\kappa + q \bar{\nu}}{2 (\bar{\alpha} + \beta ) + \bar{\pi} + \tau } \end{equation}

\begin{equation}DQ = Q (2(\epsilon + \bar{\epsilon}) + (\rho + \bar{\rho})) -Q^2 ( 2(\gamma + \bar{\gamma}) + (\mu + \bar{\mu}) )\end{equation}
\begin{equation}\Delta Q = -Q (2(\gamma + \bar{\gamma}) + (\mu + \bar{\mu}))    +q Q^2  (2(\epsilon + \bar{\epsilon}) + (\rho + \bar{\rho}))\end{equation}
\begin{equation}\delta Q = (q Q^2 - 1 )(\kappa - q\bar{\nu})\end{equation}

The factor Q proved helpful in the treatment of the IC and it is a real function since it depends only from real eigenvalues.

\subsection{Integrabillity Conditions}

The Integrability conditions come to surface by the acting of the commutation relations upon the eigenvalues. As we said in the section of the Newman Penrose Formalism the commutation relation is the Lie bracket of the basis vectors. We choose to seperate the Integrability Conditions of the Eigenvalues in two parts using the factor Q. 

\vspace{1cm}
Integrability Conditions of Eigenvalue $\lambda_0$
\begin{equation}\tag{$CR1:\lambda_0$} 2Q [D (\gamma + \bar{\gamma}) + \Delta(\epsilon+\bar{\epsilon}) + \pi \bar{\pi} - \tau \bar{\tau} ] = - [(\pi + \bar{\tau}) (q\bar{\nu} - \kappa) +  (\bar{\pi}+\tau) ( q\nu - \bar{\kappa})]\end{equation} 
\begin{multline}\tag{$CR2:\lambda_0$} Q[ 2[\delta(\epsilon + \bar{\epsilon}) - (\epsilon+\bar{\epsilon}) (\bar{\alpha} +\beta - \bar{\pi})] -[D(\bar{\pi} -\tau) -(\bar{\pi} - \tau)(\bar{\rho} +\epsilon -\bar{\epsilon})]+2\kappa(\gamma +\bar{\gamma}) -(q\bar{\nu} - \kappa)[2(\gamma+\bar{\gamma})\\
 + (\mu +\bar{\mu})] ] = D(q\bar{\nu} - \kappa) - (q\bar{\nu} - \kappa) [2\epsilon +\bar{\rho}+ \epsilon +\bar{\epsilon} +\rho+\bar{\rho}] \end{multline}
\begin{multline}\tag{$CR3:\lambda_0$}Q[2[\delta(\gamma +\bar{\gamma}) +(\gamma +\bar{\gamma})(\bar{\alpha} +\beta - \tau)] +[\Delta(\bar{\pi} -\tau) +(\bar{\pi} - \tau)(\mu - \gamma+\bar{\gamma})] -2\bar{\nu}(\epsilon +\bar{\epsilon}) -q(q\bar{\nu} - \kappa)[2(\epsilon+\bar{\epsilon})\\
 +\rho+\bar{\rho}]] = \Delta(\kappa - q\bar{\nu}) +(\kappa - q\bar{\nu})[2(\gamma +\bar{\gamma}) +(\mu +\bar{\mu}) + \mu - \gamma+\bar{\gamma}] \end{multline}
\begin{multline}\tag{$CR4:\lambda_0$} Q[\bar{\delta}(\bar{\pi}-\tau) - \delta(\pi - \bar{\tau}) - (\bar{\pi} - \tau)(\alpha - \bar{\beta}) +(\pi - \bar{\tau})(\bar{\alpha} - \beta)+2[(\epsilon + \bar{\epsilon})(\mu-\bar{\mu}) -(\gamma +\bar{\gamma})(\rho-\bar{\rho})] ]\\
 = \delta(q\nu-\bar{\kappa}) - \bar{\delta}(q\bar{\nu} - \kappa) +(q\bar{\nu}-\kappa)(\alpha-\bar{\beta}) - (q\nu - \bar{\kappa})(\bar{\alpha}-\beta)\end{multline}
\vspace{1cm}

Integrability Conditions of Eigenvalue $\lambda_1$ 

\begin{equation}\tag{$CR1:\lambda_1$}Q[ \Delta(\gamma + \bar{\gamma}) - 3(\gamma+\bar{\gamma})^2 +q[ D(\epsilon+\bar{\epsilon}) +3(\epsilon+\bar{\epsilon})^2] +\frac{q}{2}[ (\pi + \bar{\tau}) (q\bar{\nu} - \kappa) +  (\bar{\pi}+\tau) ( q\nu - \bar{\kappa})]] 
=-(\pi \bar{\pi} - \tau\bar{\tau}) 
\end{equation}
\begin{multline}\tag{$CR2:\lambda_1$} Q[ 2[ \delta(\gamma+\bar{\gamma}) -( \gamma + \bar{\gamma})(\bar{\alpha} + \beta-\bar{\pi}) ]  -q[D(q\bar{\nu}-\kappa) +(q\bar{\nu}-\kappa)(\epsilon + 3\bar{\epsilon} + \bar{\rho}) - 2\kappa(\epsilon+\bar{\epsilon})]  ] \\
= D(\bar{\pi}-\tau) - (\bar{\pi} - \tau)(\rho + 2\bar{\rho} +\epsilon-\bar{\epsilon}) - 2(\gamma + \bar{\gamma})(q\bar{\nu} - \kappa)
\end{multline}
\begin{multline}\tag{$CR3:\lambda_1$}Q[2q[\delta(\epsilon+\bar{\epsilon}) + (\epsilon +\bar{\epsilon})(\bar{\alpha}+ \beta - \tau)]+q[\Delta(q\bar{\nu}-\kappa) - (q\bar{\nu} - \kappa)(3\gamma+\bar{\gamma} -\mu) ] -2\bar{\nu}(\gamma+\bar{\gamma}) ]\\
= -[\Delta(\bar{\pi} - \tau) + (\bar{\pi}-\tau)(2\mu +\bar{\mu}-\gamma+\bar{\gamma})+2q(q\bar{\nu} - \kappa)(\epsilon+\bar{\epsilon})]
\end{multline}
\begin{multline}\tag{$CR4:\lambda_1$} Q[q[\delta(q\nu-\bar{\kappa}) - \bar{\delta}(q\bar{\nu} - \kappa) +(q\bar{\nu}-\kappa)(\alpha-\bar{\beta}) - (q\nu - \bar{\kappa})(\bar{\alpha}-\beta)] +2[q(\epsilon+\bar{\epsilon})(\rho - \bar{\rho}) - (\gamma+\bar{\gamma})(\mu-\bar{\mu})] \\
=\bar{\delta}(\bar{\pi}-\tau) - \delta(\pi - \bar{\tau}) - (\bar{\pi} - \tau)(\alpha - \bar{\beta}) +(\pi - \bar{\tau})(\bar{\alpha} - \beta) \end{multline}
\vspace{1cm}

Integrability Conditions of Eigenvalue $\lambda_2$ 
\begin{multline}\tag{$CR1:\lambda_2$} Q[ [\Delta(\mu+\bar{\mu})-(\mu+\bar{\mu})-5(\gamma+\bar{\gamma})]+q[D(\rho+\bar{\rho}) + (\rho+\bar{\rho})[(\rho+\bar{\rho})-5(\epsilon+\bar{\epsilon})]] ] \\
= \Delta(\rho+\bar{\rho}) -(\rho+\bar{\rho})(\gamma+\bar{\gamma}) + D(\mu+\bar{\mu}) +(\mu+\bar{\mu})(\epsilon+\bar{\epsilon}) \end{multline}
\begin{multline}\tag{$CR2:\lambda_2$}Q[ \delta(\mu+\bar{\mu}) -(\mu+\bar{\mu})[(\bar{\alpha}+\beta+\tau)-2\bar{\pi}] +q(\rho+\bar{\rho})(2\kappa - q\bar{\nu}) ]\\
=\delta(\rho+\bar{\rho}) -(\rho+\bar{\rho})[\bar{\alpha}+\beta+\tau-2\bar{\pi}] + (\mu+\bar{\mu})(2\kappa-q\bar{\nu})\end{multline}
\begin{multline}\tag{$CR3:\lambda_2$} qQ[\delta(\rho+\bar{\rho})+(\rho+\bar{\rho})[\bar{\alpha}+\beta+\bar{\pi}-2\tau] +(\mu+\bar{\mu})(\kappa-2q\bar{\nu})] \\
=\delta(\mu+\bar{\mu})+(\mu+\bar{\mu})[\bar{\alpha} +\beta+\bar{\pi}-2\tau]+q(\rho+\bar{\rho})(\kappa -2q\bar{\nu}) \end{multline}
\begin{multline}\tag{$CR4:\lambda_2$} Q[(\mu+\bar{\mu})(\mu-\bar{\mu}) -q(\rho+\bar{\rho})(\rho-\bar{\rho})] = (\mu-\bar{\mu})(\rho+\bar{\rho})-(\rho-\bar{\rho})(\mu+\bar{\mu})
\end{multline}

\subsection{Rotation around the null tetrad frame}
The IC along with the NPEs end up to be a fearsome system of equations. We choose to take advantage of the conformal symmetry of a rotation around one of the null tetrads $n^\mu , l^\mu$. Either of the real null tetrad remains fixed the result is the same. In our case $l$ is fixed. 

$$\tilde{\theta}^1 = e^{-a}(\theta^1+p\bar{p}\theta^2 +\bar{p}\theta^3 +p\theta^4 ) = \tilde{n}_{\mu} dx^{\mu}$$
$$\tilde{\theta}^2 = e^a \theta^2 = \tilde{l}_{\mu} dx^{\mu}$$
$$ \tilde{\theta}^3 = e^{-ib}(\theta^3 + p\theta^2)= \tilde{m}_{\mu} dx^{\mu}$$
$$\tilde{\theta}^4 = e^{ib}(\theta^4 +\bar{p}\theta^2)= \tilde{\bar{m}}_{\mu} dx^{\mu}$$
$$p\equiv c+id$$

We demand the invariant character of our Killing Tensor after the rotation. 

 $$K = \lambda_0 (\tilde{\theta}^1 \otimes \tilde{\theta}^1 +q \tilde{\theta}^2 \otimes \tilde{\theta}^2 ) +\lambda_1(\tilde{\theta}^1 \otimes \tilde{\theta}^2+\tilde{\theta}^2 \otimes \tilde{\theta}^1) + \lambda_2(\tilde{\theta}^3 \otimes \tilde{\theta}^4+\tilde{\theta}^4 \otimes \tilde{\theta}^3); \hspace{0.5cm} q= \pm 1$$

 It is easy for someone to prove that only the factor b is not annihilated, it doesn't have any contribution in the operation due to the cross terms of $\tilde{\theta}^3 , \tilde{\theta}^4$. The rotation applied also in the spin coefficients.
 $$\sigma=0=\lambda$$
 $$\tilde{\rho} = \rho \hspace{1cm}  \tilde{\mu} = \mu$$
 $$\tilde{\kappa} = e^{ib} \kappa \hspace{1cm} \tilde{\tau} = e^{ib} \tau$$    
 $$\tilde{\pi} = e^{-ib} \pi \hspace{1cm} \tilde{\nu}=e^{-ib} \nu $$ 
 $$\tilde{\alpha} = e^{-ib}(\alpha +\frac{\bar{\delta}{(ib)}}{2}) \hspace{1cm} \tilde{\beta} = e^{ib}(\beta+\frac{\delta(ib)}{2})$$
  $$\tilde{\epsilon} = \epsilon +\frac{D(ib)}{2} \hspace{1cm} \tilde{\gamma} = \gamma+\frac{\Delta(ib)}{2}$$

Now trying to assimilate the information we set the four last tilded coefficients equal to zero. Then we feed the four last relations into the CR resulting the $\bf{Key}$ $\bf{relations}$ which unfold the family branches of the solution.    
 
 \begin{equation}\tag{i} \Psi_2 - \Lambda = \kappa \nu - \tau \pi \end{equation}
 \begin{equation}\tag{ii}\Psi_1 = \kappa \mu\end{equation}
 \begin{equation}\tag{iii}\Psi_2 - \Lambda = \mu \rho\end{equation}
\begin{equation}\tag{iv}\mu \tau = 0\end{equation}

 The rotation provides us with these useful relations which connect the spin coefficients along with the Weyl components. This result depends mainly in the form of the Killing Tensor since we demand by the rotation to preserves the Killing tensor invariant. Essentially the lack either of $\lambda_0$ or $\lambda_7$ supplies us with the key relations. 

 \textbf{Remark} :We conclude that we get additional information of the conformal symmetry of the rotation only if one the non-diagonal elements of the Killing Tensor are absent. Regarding the latter, we could get corresponding results if we annihilated $\lambda_0$ at the beginning instead of $\lambda_7$.  In case we choose to deal with the full expression for Killing Tensor we wouldn't get further information based on the implying symmetry. 
 
 The latter is a remarkable assistance in the pursuit of the solution. We present the classes of solutions due to the key relations. Implying the relation $\mu \tau = 0 $ to NPEs, IC, BI provide us with the three main classes of our solution. These classes appointed by the three branches of the latter relation. Also, we aim to study solutions in which only one of the eigenvalues of Killing tensor $ \lambda_1, \lambda_2, \lambda_3$ is allowed to be constant.
 
It would be helpful for the reader if in this point we array the most useful NPEs, the $CR4:\lambda_2$ along with the key relations, besides we mainly use these relations     
in order to determine the classes of the solution. 

\begin{equation}\tag{b} \delta \kappa = \kappa ( \tau - \bar{\pi} +2(\bar{\alpha} +\beta) - (\bar{\alpha} - \beta) ) - \Psi _o  \end{equation}
\begin{equation}\tag{g} \bar{\delta} \pi = - \pi(\pi + \alpha - \bar{\beta}) + \nu\bar{\kappa} \end{equation}
\begin{equation}\tag{p} \delta \tau = \tau (\tau - \bar{\alpha} + \beta) -\bar{\nu} \kappa  \end{equation}
\begin{equation}\tag{k} \delta\rho = \rho(\bar{\alpha} + \beta) +\tau(\rho-\bar{\rho})+ \kappa(\mu-\bar{\mu}) - \Psi_1 \end{equation}
\begin{equation}\tag{m} \bar{\delta} \mu = -\mu (\alpha +\bar{\beta}) -\pi (\mu- \bar{\mu}) - \nu (\rho-\bar{\rho}) + \Psi_3 \end{equation}
\begin{equation}\tag{j} \bar{\delta}\nu = - \nu(2(\alpha +\bar{\beta} ) + (\alpha - \bar{\beta})  + \pi - \bar{\tau}  ) + \Psi_4\end{equation}
 
 \begin{multline}\tag{$CR4:\lambda_2$} Q[(\mu+\bar{\mu})(\mu-\bar{\mu}) -q(\rho+\bar{\rho})(\rho-\bar{\rho})] = (\mu-\bar{\mu})(\rho+\bar{\rho})-(\rho-\bar{\rho})(\mu+\bar{\mu})
\end{multline}

 \subsection{\it{Class I : $\mu = 0$}}
 
  The implication of the annihilation of $\mu$ to key relations to IC, to NPE and to BI emerged our classes.
  The Key relations now take the form with $\Psi_2 =\Lambda = constant$ 
   \begin{equation}\tag{i} \Psi_2 - \Lambda = 0 = \kappa \nu - \tau \pi \end{equation}
 \begin{equation}\tag{ii}\Psi_1 = 0\end{equation}
 \begin{equation}\tag{iii}\Psi_2 - \Lambda = 0\end{equation}
\begin{equation}\tag{iv}\mu = 0\end{equation}

Consider that $\mu=0$ the relation $CR4:\lambda_2 $ gives $(\rho +\bar{\rho})(\rho-\bar{\rho})=0$. According to IC (17)-(19) the derivative of the eigenvalue $\lambda_2$ depends only from the real parts of $\mu,\rho$ so our priority is to avoid the extinction of $\rho +\bar{\rho} = 0$ because the choice $\rho-\bar{\rho} =0$ guarantees that the eigenvalue $\lambda_2$ is not a constant since we already have $\mu = 0$. Unavoidably though the BI (II) giving birth to solutions where derivative of $\lambda_2$ vanishes. Anyway from NPE (m) we take $\Psi_3 = \nu (\rho - \bar{\rho}) = 0 $. Hence the BI takes the form
  
  The Bianchi Identities

\begin{equation}\tag{I} \bar{\delta} \Psi_0  = (4\alpha - \pi )\Psi_0  +3\kappa \Psi_2\end{equation}
\begin{equation}\tag{II} 0= (\rho +\bar{\rho}) \Psi_2 \end{equation}
\begin{equation}\tag{III}  0= -3\pi \Psi_2 +\kappa\Psi_4\end{equation}
\begin{equation}\tag{IV}  D\Psi_4 = -4\epsilon \Psi_4 \end{equation} 
\begin{equation}\tag{V} \Delta \Psi_0  = 4\gamma  \Psi_0 \end{equation}
\begin{equation}\tag{VI} \Delta \Psi_1  = \nu \Psi_0  -3\tau\Psi_2\end{equation}
\begin{equation}\tag{VIII} -\delta \Psi_4 = 3\nu \Psi_2 +(4\beta-\tau)\Psi_4\end{equation} 

 In the following tables every column represents different solutions according to differenct choices which are at our disposal and which are ordered by the key relations and specifically by the BI (II). The collumns of the tables contain the main charasteristics of our solutions. The second and third collumn of Table 1 distinguished by the different choices that takes place due to the BI (III) and (VI).
\begin{table}
		\centering 
			\caption{ $\rho -\bar{\rho} =  0 = \rho+\bar{\rho}$ }
		\begin{tabular}{ c c c }
			\toprule
			$\rho= 0 \neq \Psi_2$ & $\rho + \bar{\rho} = 0 = \Psi_2$ & $\rho + \bar{\rho} = 0 = \Psi_2$  \\
			Type D& Type N& Type N  \\
			\toprule
$ \kappa \nu = \pi \tau$   & $ \nu =0 =  \pi \tau  $ & $  \kappa =0 =  \pi \tau$ \\
$\Psi_0 \Psi_4 = {9\Psi_2}^2$ & $\Psi_0\neq0 $    &$\Psi_4\neq0 $\\
 $\Psi_1= \Psi_3 = 0$  &$ \Psi_1= \Psi_2 = \Psi_3 = \Psi_4 = 0$  & $\Psi_0 = \Psi_1= \Psi_2 =\Psi_3 = 0$ \\
			\midrule
 $d\lambda_2 = 0$     & $d\lambda_2 = 0$ &    $d\lambda_2 = 0$ \\ 
			\bottomrule
		\end{tabular}
	\end{table}

The other choice where $\rho+\bar{\rho}\neq0=\Psi_2$ yields Type N solutions where the eigenvalue $\lambda_2$ is not a constant. The combinations of BI (III) and (VI) determines which of the Weyl component is zero.

\begin{table}
		\centering 
			\caption{ $\rho -\bar{\rho} =  0 \neq \rho+\bar{\rho}$ }
		\begin{tabular}{ c c c c c c }
			\toprule
			Type N & Type N & Type N &Type N &Type N & Type N   \\ 
			\toprule
   $\nu =0= \tau$  &  $\nu =0 =\pi$ &  $\kappa =0 =\tau $ & $  \kappa =0 =\pi $ &  $ \nu = \pi= \tau=0$ & $ \kappa = \pi =\tau = 0$  \\ 
  $\Psi_0\neq 0$            & $\Psi_4 \neq 0$& $\Psi_0\neq 0$            & $\Psi_4 \neq 0$&  $\Psi_0\neq 0$     & $\Psi_4 \neq 0$  \\ 
 
  		\midrule
 $d\lambda_2 \neq 0$     &  $d\lambda_2 \neq 0$  & $d\lambda_2 \neq 0$     & $d\lambda_2 \neq 0$     & $d\lambda_2 \neq 0$    & $d\lambda_2 \neq 0$   \\ 
  		
			\bottomrule
		\end{tabular}
	\end{table}

 \subsection{\it{Class II: $\mu = 0 = \tau$}}
 As before the NPE (m) gives us $\Psi_3 = \nu (\rho - \bar{\rho}) = 0$. From key relation (ii) we take again $\Psi_1 = 0$ and from (i) $\kappa \nu = 0$. Now it is obvious that the Class II consist a subset of the Class I. These solutions are similar to the Type N solutions  with $\tau = 0$.
 
 \subsection{\it{Class III: $\tau = 0$}}
In this class we encountered new algebraically special solutions. NPE (p) for $\tau =0$ yields $\bar{\kappa} \nu = 0$. Due to the latter the key relations rewritten as follows.
 \begin{equation}\tag{i} \Psi_2 - \Lambda =0  = \kappa \nu  \end{equation}
 \begin{equation}\tag{ii}\Psi_1 = \kappa \mu\end{equation}
 \begin{equation}\tag{iii} \Psi_2 - \Lambda = 0 = \mu \rho \end{equation}
\begin{equation}\tag{iv} \tau = 0\end{equation}

The branch where $\mu = 0$ is already known from the Class I, II which are Type N solutions with a further simplification $\tau=0$. On the other hand, the case of $\mu \neq  0 = \rho $ giving birth to few solutions which are worth to be mentioned. 
Crucial role to this Class plays also the $CR4:\lambda_2$ since the annihilation of $\rho$ implies $(\mu +\bar{\mu})(\mu - \bar{\mu}) = 0$. The constraint $\mu \neq 0$ leads us to two separate solutions for the case $\kappa = 0 \neq \nu$ which are both of Type III. The other case where both $\kappa$ and $\nu$ are zero concerns two solutions where only $\Psi_3$ is not equal to zero determining that the type of  the solutions are also Type III.

The last branch contains the case $\kappa \neq 0 = \nu$, it should be noted that NPE (k) with $\rho = 0$ gives us the form of $\Psi_1$, which resulting to the annihilation of the real part of $\mu$ and in the same time dictates that $d\lambda_2 = 0$ since we know that the derivatives of $\lambda_2$ depends only from the real part of $\mu$ and $\rho$. Along these lines we gain the following. 

\begin{equation}\tag{k} \Psi_1 = \kappa (\mu - \bar{\mu}) \end{equation}

Finally the Class III is presented in the following table. In all cases $\Psi_2 = \Lambda=0$
resulted by NPE (q) for $\rho = \tau = 0$ while in these Type III cases $\Psi_1 = 0$.
\begin{table}
		\centering 
			\caption{ $\mu \neq  0 = \rho$ }
		\begin{tabular}{ c c  }
			\toprule
			$\mu - \bar{\mu}= 0 $ & $\mu + \bar{\mu} = 0$   \\
			Type III& Type III \\
			\toprule
$ \kappa=0\neq \nu $   & $ \kappa =0 \neq \nu  $  \\
$\Psi_3\neq 0 \neq \Psi_4 $ & $\Psi_3\neq0 \neq \Psi_4$  \\
			\midrule
 $d\lambda_2 \neq 0$     &  $d\lambda_2 = 0$  \\ 
			\bottomrule
		\end{tabular}
	\end{table}
Finally we reach to the point to present the following theorem. 
\textbf{Theorem}\textit{ All Petrov Types except Type I and Type II admit both $K^2_{\mu \nu}, K^3_{\mu \nu}$ Canonical Forms of Killing Tensor. }

\section{Considerations and Solution}
In this work we are focused on the Type D solution of the Class I for the $q=+1$ case and the following theorem is going to be proven.

\textbf{theorem}\textit{Carter's case [$\mathcal{D}$] admits the existence of 2nd Canonical Form of a Killing Tensor $K^2_{\mu \nu}$.} 

Our solution is of Type D and the components of Weyl tensor are connected by the following relation $\Psi_0 \Psi_4 = {9\Psi_2}^2$ with $\Psi_2 = \Lambda $. Generally speaking Type D spacetimes have only one non-zero Weyl component, the $\Psi_2$. Although there are also two other versions. The first version characterized by the relation $3\Psi_2 \Psi_4 = 2\Psi_3^2$ where $\Psi_0 = \Psi_1 = 0$ and the second version is like ours where $\Psi_0 \Psi_4 = {9\Psi_2}^2$ with $\Psi_1 = \Psi_3 = 0$  \cite{griffiths2009exact}. At last all these versions are equivalent and could be obtained with classes of rotations, Chandrasekhar and Xanthopoulos \cite{chandrasekhar1986new} prove that in a chosen null-tetrad frame the type D character of our case could be obtained with two classes of rotation around the two vectors l,n of the null-tetrad frame.

In this section we demonstrate the useful relations that determine our solution. Also taking advantage by the rotation with $l$ fixed we obtained the Key relations. \textbf{This is the maximal utilization of symmetry that one could gain from a rotation around the null tetrad frame with the 2nd Canonical Jordan form of the Killing Tensor with $\lambda_7$}. 

$$\sigma = 0 = \lambda$$
$$ \mu = 0 = \rho $$
$$\Psi_0 \Psi_4 = 9 \Psi^2_2$$
$$\Psi_2 = \Lambda = Constant$$
$$\Psi_1 = 0 = \Psi_3$$
$$\kappa \nu = \tau \pi$$
 $$\bar{\alpha} +\beta = 0$$
$$ \epsilon +\bar{\epsilon} = 0$$
$$\gamma+\bar{\gamma} = 0$$

The NPEs, BI, IC with the substitution of the upward relations are given by the following.
 Newman Penrose Equations

\begin{equation}\tag{a}  \bar{\delta} \kappa =  \bar{\kappa} \tau + \kappa ( (\alpha - \bar{\beta}) - \pi) \end{equation}
\begin{equation}\tag{b} \delta \kappa = \kappa ( \tau - \bar{\pi}  - (\bar{\alpha} - \beta) ) - \Psi _o  \end{equation}
\begin{equation}\tag{c} D\tau = \Delta \kappa + \tau(\epsilon - \bar{\epsilon})    -2\kappa \gamma  \end{equation}
\begin{equation}\tag{i} D\nu - \Delta \pi = \pi(\gamma - \bar{\gamma}) -2\nu \epsilon   \end{equation}
\begin{equation}\tag{g} \bar{\delta} \pi = - \pi(\pi + \alpha - \bar{\beta}) + \nu\bar{\kappa} \end{equation}
\begin{equation}\tag{p} \delta \tau = \tau (\tau - \bar{\alpha} + \beta) -\bar{\nu} \kappa  \end{equation}
\begin{equation}\tag{h} - \delta\pi =  \pi(\bar{\pi} - \bar{\alpha} +\beta) - \kappa \nu + \Psi_2 + 2\Lambda\end{equation}
\begin{equation}\tag{n} \delta\nu  =  - \bar{\nu} \pi + \nu (\tau  +(\bar{\alpha} -\beta) )\end{equation}
\begin{equation}\tag{q} - \bar{\delta} \tau = - \tau(\bar{\tau} + \alpha - \bar{\beta}) + \nu \kappa  - \Psi_2 - 2\Lambda\end{equation}
\begin{equation}\tag{k}  \Psi_1 = 0 \end{equation}
\begin{equation}\tag{m}  \Psi_3 = 0 \end{equation}
\begin{equation}\tag{d} D\alpha - \bar{\delta} \epsilon = \alpha( \bar{\epsilon} -2\epsilon) - \bar{\beta}\epsilon - \bar{\kappa}\gamma + \pi \epsilon \end{equation} 
\begin{equation}\tag{e} D \beta - \delta{\epsilon} = - \beta  \bar{\epsilon}-\kappa \gamma -\epsilon(\bar{\alpha} - \bar{\pi}) \end{equation} 
\begin{equation}\tag{r} \Delta \alpha - \bar{\delta}\gamma = \nu\epsilon +\alpha \bar{\gamma}  +\gamma (\bar{\beta}- \bar{\tau})  \end{equation}
\begin{equation}\tag{o} -\Delta \beta + \delta \gamma = \gamma\tau - \epsilon \bar{\nu} - \beta( \gamma - \bar{\gamma} )\end{equation}
\begin{equation}\tag{l} \delta \alpha - \bar{\delta}\beta = \alpha (\bar{\alpha} - \beta) - \beta(\alpha - \bar{\beta}) \end{equation}
\begin{equation}\tag{f} D\gamma - \Delta \epsilon =  \alpha(\tau + \bar{\pi}) + \beta( \bar{\tau} + \pi) \end{equation}
\begin{equation}\tag{j} \bar{\delta}\nu = - \nu( \alpha - \bar{\beta} + \pi - \bar{\tau}  ) + \Psi_4\end{equation}
\vspace{0.5cm}
The Bianchi Identities

\begin{equation}\tag{I} \bar{\delta} \Psi_0  = (4\alpha - \pi )\Psi_0   +3\kappa \Psi_2\end{equation}
\begin{equation}\tag{II} 0=0 \end{equation}
\begin{equation}\tag{III} 3\pi \Psi_2 =\kappa\Psi_4\end{equation}
\begin{equation}\tag{IV} D\Psi_4 = - 4\epsilon \Psi_4 \end{equation} 
\begin{equation}\tag{V} \Delta \Psi_0 = 4\gamma  \Psi_0 \end{equation}
\begin{equation}\tag{VI} \nu \Psi_0  = 3\tau\Psi_2\end{equation}
\begin{equation}\tag{VII} 0=0\end{equation} 
\begin{equation}\tag{VIII} - \delta \Psi_4 = 3\nu \Psi_2 +(4\beta-\tau)\Psi_4\end{equation}

\vspace{1cm}
Integrability Conditions of Eigenvalue $\lambda_0$
\begin{equation}\tag{$CR1:\lambda_0$} 2Q [ \pi \bar{\pi} - \tau \bar{\tau} ] =  (\pi + \bar{\tau}) ( \kappa-q\bar{\nu}) +  (\bar{\pi}+\tau) (  \bar{\kappa}-q\nu)\end{equation} 
\begin{equation}\tag{$CR2:\lambda_0$} Q [D(\bar{\pi} -\tau) -(\bar{\pi} - \tau)(\epsilon -\bar{\epsilon})] = D(\kappa - q\bar{\nu}) -(\epsilon -\bar{\epsilon}) ( \kappa - q\bar{\nu})  \end{equation}
\begin{equation}\tag{$CR3:\lambda_0$}Q [\Delta(\bar{\pi} - \tau) -(\gamma -\bar{\gamma}) (\bar{\pi}-\tau) ] = \Delta(\kappa-q\bar{\nu}) - (\gamma-\bar{\gamma})( \kappa -q\bar{\nu}) \end{equation}
\begin{multline}\tag{$CR4:\lambda_0$} Q[\bar{\delta}(\bar{\pi}-\tau) - \delta(\pi - \bar{\tau}) - (\bar{\pi} - \tau)(\alpha - \bar{\beta}) +(\pi - \bar{\tau})(\bar{\alpha} - \beta) ]\\
 = \delta(q\nu-\bar{\kappa}) - \bar{\delta}(q\bar{\nu} - \kappa) +(q\bar{\nu}-\kappa)(\alpha-\bar{\beta}) - (q\nu - \bar{\kappa})(\bar{\alpha}-\beta)\end{multline}

Integrability Conditions of Eigenvalue $\lambda_1$ 

\begin{equation}\tag{$CR1:\lambda_1$}Q\frac{q}{2}[ (\pi + \bar{\tau}) ( \kappa -q\bar{\nu}) +  (\bar{\pi}+\tau) (  \bar{\kappa} - q\nu)]=(\pi \bar{\pi} - \tau\bar{\tau}) 
\end{equation}
\begin{equation}\tag{$CR2:\lambda_1$} qQ[D(\kappa-q\bar{\nu}) -(\epsilon -\bar{\epsilon})(\kappa -q\bar{\nu}) ] = D(\bar{\pi}-\tau) - (\bar{\pi} - \tau)(\epsilon-\bar{\epsilon}) \end{equation}
\begin{equation}\tag{$CR3:\lambda_1$}qQ[\Delta(\kappa-q\bar{\nu}) - (\gamma-\bar{\gamma})( \kappa -q\bar{\nu})]= \Delta(\bar{\pi} - \tau) -(\gamma -\bar{\gamma}) (\bar{\pi}-\tau)\end{equation}
\begin{multline}\tag{$CR4:\lambda_1$} qQ[\delta(q\nu -\bar{\kappa}) - \bar{\delta}(q\bar{\nu} - \kappa) +(q\bar{\nu}-\kappa)(\alpha-\bar{\beta}) - (q\nu - \bar{\kappa})(\bar{\alpha}-\beta)]  \\
=\bar{\delta}(\bar{\pi}-\tau) - \delta(\pi - \bar{\tau}) - (\bar{\pi} - \tau)(\alpha - \bar{\beta}) +(\pi - \bar{\tau})(\bar{\alpha} - \beta) \end{multline}

In this point should be noted that due to the annihilation of $\mu$ and $\rho$ coefficients the eigenvalue $\lambda_2$ is constant. So we don't take any information from its commutation relations. Also it is obvious that from the corresponding commutation relations we take the same information since the condition $qQ^2-1=0$ does not valid for both $K^2_{\mu \nu}$ and $K^3_{\mu \nu}$. Hence the CR are summarized downward.    

\begin{equation}\bar\pi \pi - \tau \bar{\tau} = 0 \end{equation}
\begin{equation} \bar{\kappa} \kappa - \bar{\nu} \nu =0  \end{equation}
\begin{equation}D(\kappa-q\bar{\nu}) -(\epsilon -\bar{\epsilon})(\kappa -q\bar{\nu})=0 \end{equation}
\begin{equation} D(\bar{\pi}-\tau) - (\epsilon-\bar{\epsilon})(\bar{\pi} - \tau)=0 \end{equation}
\begin{equation}\Delta(\kappa-q\bar{\nu}) - (\gamma-\bar{\gamma})( \kappa -q\bar{\nu})=0 \end{equation}
\begin{equation}\Delta(\bar{\pi} - \tau) -(\gamma -\bar{\gamma}) (\bar{\pi}-\tau) = 0 \end{equation}
\begin{equation} \delta(q\nu-\bar{\kappa}) - \bar{\delta}(q\bar{\nu} - \kappa) +(q\bar{\nu}-\kappa)(\alpha-\bar{\beta}) - (q\nu - \bar{\kappa})(\bar{\alpha}-\beta) =0 \end{equation}
\begin{equation}\bar{\delta}(\bar{\pi}-\tau) - \delta(\pi - \bar{\tau}) - (\bar{\pi} - \tau)(\alpha - \bar{\beta}) +(\pi - \bar{\tau})(\bar{\alpha} - \beta)= 0\end{equation}

\subsection{Frobenius Theorem}
Considering the relations (11),(12),(13) we can make a suitable choice for our spin coefficients. One possible solution is the following. 

\begin{equation}\bar{\pi}+\tau = 0 \end{equation}
\begin{equation}\kappa +q\bar{\nu} = 0\end{equation}
 \begin{equation}\bar{\alpha} +\beta = 0 \end{equation}

This solution regards only the form $K^2_{\mu \nu}$ since the equation $\kappa \nu =\pi \tau$ after the substitution of (32),(33) dictates $q=+1$.  We shall now proceed to the implication of the Frobenius Integrability Theorem.

The Cartan's structure is

\begin{equation}d\theta^1 = -\bar{\pi}\theta^1\wedge\theta^3 - \pi \theta^1\wedge\theta^4 -\bar{\nu}\theta^2\wedge\theta^3 -\nu\theta^2\wedge\theta^4\end{equation}
\begin{equation}d\theta^2 = \kappa\theta^1\wedge\theta^3 +\bar{\kappa} \theta^1\wedge\theta^4 +\tau\theta^2\wedge\theta^3 +\bar{\tau}\theta^2\wedge\theta^4\end{equation}
\begin{equation}d\theta^3 = -(\epsilon-\bar{\epsilon})\theta^1\wedge\theta^3 -(\gamma-\bar{\gamma})\theta^2\wedge\theta^3 +(\alpha-\bar{\beta})\theta^3\wedge\theta^4\end{equation}
\begin{equation}d\theta^4 = -(\epsilon-\bar{\epsilon})\theta^1\wedge\theta^4 -(\gamma-\bar{\gamma})\theta^2\wedge\theta^4 -(\bar{\alpha}-\beta)\theta^3\wedge\theta^4\end{equation}

It follows that

\begin{equation}d\theta^1 \wedge\theta^1 \wedge\theta^2 =0  \end{equation}
\begin{equation}d\theta^2 \wedge\theta^1 \wedge\theta^2 =0  \end{equation}
\begin{equation}d(\theta^3-\theta^4)  \wedge(\theta^3-\theta^4) \wedge(\theta^3 +\theta^4) =0  \end{equation}
\begin{equation}d(\theta^3+\theta^4)  \wedge(\theta^3-\theta^4) \wedge(\theta^3 +\theta^4) =0  \end{equation}

which on account of Frobenius Integrability Theorem implies the existence of a local coordinate system of (t,z,x,y) such that

\begin{equation}\theta^1 = (L-N)dt +(M-P)dz\end{equation}
\begin{equation}\theta^2 = (L+N)dt +(M+P)dz\end{equation}
\begin{equation}\theta^3 = Sdx +i Rdy\end{equation}
\begin{equation}\theta^4 = Sdx -i Rdy\end{equation}

where L,N,M,P,S,R are real valued functions of (t,z,x,y) \footnote{In this point it should be noted that the downward indices denotes the derivation in respect to coordinates.}. Now if one replaces the differential forms in (35)-(38) by their values (43)-(46) and equates the corresponding coefficients of the differentials it follows that
 
 \begin{equation}R_t = R_z = S_t = S_z = 0  \Rightarrow \gamma-\bar{\gamma}  = \epsilon-\bar{\epsilon} = 0\end{equation}
\begin{equation} M_t = L_z\end{equation}
\begin{equation} P_t = N_z\end{equation}
\begin{equation} M_x L-L_x M = 0 =M_y L-L_y M = 0\end{equation}
\begin{equation} P_x N-N_x P = 0 =P_y N-N_y P = 0\end{equation}
\begin{equation} \bar{\pi}=-\tau = \frac{\delta Z}{2Z} \end{equation}
\begin{equation} \kappa = -\bar{\nu} = \frac{(M_x N - N_x M) +(P_x L - L_x P)}{4ZS} -i\frac{(M_y N - N_y M) +(P_y L - L_y P)}{4ZR} \end{equation}
\begin{equation}2\alpha = \alpha -\bar{\beta} = \frac{-1}{2}(\frac{(\delta +\bar{\delta})R}{R} - \frac{(\delta - \bar{\delta})S}{S}) \end{equation}
\begin{equation}Z\equiv PL-MN\end{equation}

Now taking advantage of relations (50),(51)

\begin{equation} L=A(t,z)M \end{equation}
\begin{equation} N=B(t,z)P \end{equation}

 and substitutes them into (48)-(55) gives the following.

\begin{equation} M_t = (AM)_z\end{equation}
\begin{equation} P_t = (B P)_z\end{equation}
\begin{equation} \bar{\pi}=-\tau = \frac{\delta (PM)}{2PM} \end{equation}
\begin{equation} \kappa = -\bar{\nu} = \frac{1}{2}(\frac{\delta P}{P} - \frac{\delta M}{M}) \end{equation}
\begin{equation}2\alpha = \alpha -\bar{\beta} = -\frac{1}{2}(\frac{(\delta +\bar{\delta})R}{R} - \frac{(\delta - \bar{\delta})S}{S}) \end{equation}
\begin{equation}Z\equiv PL-MN = (A-B)PM\end{equation}

The products of the implication of the Frobenius theorem have a great impact in the NPEs, BI, IC.  

Newman Penrose Equations
\begin{equation}\tag{a}  \bar{\delta} \kappa =  \bar{\kappa} \tau + \kappa ( (\alpha - \bar{\beta}) - \pi) \end{equation}
\begin{equation}\tag{b} \delta \kappa = \kappa ( \tau - \bar{\pi}  - (\bar{\alpha} - \beta) ) - \Psi _o  \end{equation}
\begin{equation}\tag{g} \bar{\delta} \pi = - \pi(\pi + \alpha - \bar{\beta}) + \nu\bar{\kappa} \end{equation}
\begin{equation}\tag{p} \delta \tau = \tau (\tau - \bar{\alpha} + \beta) -\bar{\nu} \kappa  \end{equation}
\begin{equation}\tag{h} - \delta\pi =  \pi(\bar{\pi} - \bar{\alpha} +\beta) - \kappa \nu + \Psi_2 + 2\Lambda\end{equation}
\begin{equation}\tag{n} \delta\nu  =  - \bar{\nu} \pi + \nu (\tau  +(\bar{\alpha} -\beta) )\end{equation}
\begin{equation}\tag{q} - \bar{\delta} \tau = - \tau(\bar{\tau} + \alpha - \bar{\beta}) + \nu \kappa  - \Psi_2 - 2\Lambda\end{equation}
\begin{equation}\tag{d} D\alpha = D \beta = 0 \end{equation} 
\begin{equation}\tag{r}\Delta \alpha = \Delta \beta = 0 \end{equation}
\begin{equation}\tag{l} \delta \alpha - \bar{\delta}\beta = \alpha (\bar{\alpha} - \beta) - \beta(\alpha - \bar{\beta}) \end{equation}
\begin{equation}\tag{j} \bar{\delta}\nu = - \nu( \alpha - \bar{\beta} + \pi - \bar{\tau}  ) + \Psi_4\end{equation}
\vspace{0.5cm}

Bianchi Identities require a reformation in order to be functionable. Along these lines it is easy to correlate $\Psi_0$ with $\Psi_4$ combining BI (III) with BI (VI). Next, we aim to abolish $\Psi_4$ by our relations. Hence we multiply BI (IV) with $\pi$ then with the usage of $\kappa \nu = \pi \tau$ we get

\begin{equation}\tag{VI} 3\kappa \Psi_2 = \pi \Psi_0 \end{equation}

The latter combined with BI (I) gives 
 
\begin{equation}\tag{I} \bar{\delta} \Psi_0  = 4\alpha \Psi_0 \end{equation}
\begin{equation}\tag{IV} D\Psi_0 = 0 \end{equation} 
\begin{equation}\tag{V} \Delta \Psi_0 =0 \end{equation}

Where the relations between the Weyl components are given by
 
\begin{equation} \Psi_0 = \Psi^*_4 \end{equation}
\begin{equation} \Psi_4 \Psi^*_4 = \Psi_0 \Psi^*_0 = 9 \Lambda^2 \end{equation}

At last the Integrability Conditions resulted to the following where (68) and (69) are relations where can be obtained by (a), (b). 

\begin{equation}D \kappa = \Delta \kappa =D\nu = \Delta \nu = 0 \end{equation}
\begin{equation} D \pi = \Delta \pi = D \tau = \Delta \tau =0  \end{equation}
\begin{equation} \delta\bar{\kappa} - \bar{\delta}\kappa=\kappa(\alpha-\bar{\beta})  - \bar{\kappa}(\bar{\alpha}-\beta)  \end{equation}
\begin{equation}\bar{\delta} \bar{\pi}- \delta \pi = \bar{\pi} (\alpha - \bar{\beta}) -\pi(\bar{\alpha} - \beta)\end{equation}

The relations (d),(r),(66)(67) make clear that our metric doesn't depend from t,z since every spin coefficient is annihilated both by $D, \Delta$. As we know, the type D solutions  admit a Riemannian-Maxwellian invertible structure and hence there is an invertible abelian two parameter isometry group. This have been proved by \cite{debever1981orthogonal}, \cite{debever1979riemannian}. Considering that our two commutative Killing vectors are $\partial_t,\partial_z$ or a combination of these two then our equations can be expressed as follows.  

\vspace{1cm}
Newman Penrose Equations
\begin{equation}(\delta +\bar{\delta})( \pi +\bar{\pi}) = -(\pi +\bar{\pi})^2 -(\kappa +\bar{\kappa})^2  - (\pi-\bar{\pi})[(\alpha - \bar{\beta})-(\bar{\alpha}-\beta)]- 6\Psi_2 \end{equation}
\begin{equation}(\delta -\bar{\delta})( \pi -\bar{\pi}) = (\pi -\bar{\pi})^2 +(\kappa -\bar{\kappa})^2  + (\pi+\bar{\pi})[(\alpha - \bar{\beta})+(\bar{\alpha}-\beta)]- 6\Psi_2  \end{equation}
\begin{equation}(\delta +\bar{\delta})( \pi -\bar{\pi}) = -(\pi - \bar{\pi})(\pi +\bar{\pi}) +(\kappa +\bar{\kappa})(\kappa-\bar{\kappa})  - (\pi+\bar{\pi})[(\alpha - \bar{\beta})-(\bar{\alpha}-\beta)]  \end{equation}
\begin{equation}(\delta -\bar{\delta})( \pi +\bar{\pi}) = (\pi - \bar{\pi})(\pi +\bar{\pi}) -(\kappa +\bar{\kappa})(\kappa-\bar{\kappa})  + (\pi-\bar{\pi})[(\alpha - \bar{\beta})+(\bar{\alpha}-\beta)]  \end{equation}
\begin{equation} (\delta+\bar{\delta})(\kappa +\bar{\kappa}) = -2(\pi+\bar{\pi})(\kappa+\bar{\kappa}) +(\kappa-\bar{\kappa})[(\alpha-\bar{\beta})-(\bar{\alpha} - \beta)]-(\Psi_0 +\Psi_0^*)\end{equation}
\begin{equation} (\delta-\bar{\delta})(\kappa -\bar{\kappa}) = 2(\pi-\bar{\pi})(\kappa-\bar{\kappa}) +(\kappa+\bar{\kappa})[(\alpha-\bar{\beta})+(\bar{\alpha} - \beta)]-(\Psi_0 +\Psi_0^*) \end{equation}
\begin{equation}  (\delta-\bar{\delta})(\kappa +\bar{\kappa}) = -(\pi+\bar{\pi})(\kappa-\bar{\kappa}) +(\pi-\bar{\pi})(\kappa+\bar{\kappa}) -(\kappa-\bar{\kappa})[(\alpha-\bar{\beta})+(\bar{\alpha} - \beta)]-(\Psi_0 -\Psi_0^*)\end{equation}
\begin{equation}(\delta+\bar{\delta})(\kappa -\bar{\kappa}) = -(\pi+\bar{\pi})(\kappa-\bar{\kappa}) +(\pi-\bar{\pi})(\kappa+\bar{\kappa}) +(\kappa+\bar{\kappa})[(\alpha-\bar{\beta})-(\bar{\alpha} - \beta)]-(\Psi_0 -\Psi_0^*) \end{equation}
\begin{equation}\delta(\alpha -\bar{\beta}) +\bar{\delta}(\bar{\alpha} -\beta) =2(\alpha -\bar{\beta})(\bar{\alpha} - \beta) \end{equation}
\vspace{0.5cm}

Bianchi Identities
\begin{equation}\tag{I} \bar{\delta} \Psi_0  = 4\alpha \Psi_0 \end{equation}
\begin{equation}\tag{VI} 3\kappa \Psi_2 = \pi \Psi_0 \end{equation}

 Using the relations for spin coefficients 

\begin{equation}  \bar{\pi}=-\tau = \frac{\delta (PM)}{2PM} \end{equation}
\begin{equation} \kappa = -\bar{\nu} = \frac{1}{2}(\frac{\delta P}{P} - \frac{\delta M}{M}) \end{equation}
\begin{equation}2\alpha = \alpha -\bar{\beta} = -\frac{1}{2}(\frac{(\delta +\bar{\delta})R}{R} - \frac{(\delta - \bar{\delta})S}{S}) \end{equation}
\begin{equation}Z\equiv PL-MN = (A-B)PM\end{equation} 

\begin{equation} (\delta +\bar{\delta}) = \frac{\partial_x}{S} \end{equation}
\begin{equation} (\delta -\bar{\delta}) = (-i)\frac{\partial_y}{R} \end{equation}

our NPEs are listed below.

\begin{equation} 12\Psi_2 = -\frac{1}{PR}  \left[ \left[\frac{P_y}{R}\right]_y +\frac{R_x}{S}\frac{P_x}{S} \right] - \frac{1}{MR} \left[ \left[ \frac{M_y}{R}\right]_y + \frac{R_x}{S}\frac{M_x}{S}  \right] \end{equation}

\begin{equation} 12\Psi_2 = -\frac{1}{PS}  \left[ \left[\frac{P_x}{S}\right]_x +\frac{S_y}{R}\frac{P_y}{R} \right] - \frac{1}{MS} \left[ \left[ \frac{M_x}{S}\right]_x + \frac{S_y}{R}\frac{M_y}{R}  \right] \end{equation}

\begin{equation} 2(\Psi_0 +\Psi_0^*) = \frac{1}{PR}  \left[ \left[\frac{P_y}{R}\right]_y +\frac{R_x}{S}\frac{P_x}{S} \right] - \frac{1}{MR} \left[ \left[ \frac{M_y}{R}\right]_y + \frac{R_x}{S}\frac{M_x}{S}  \right] \end{equation}

\begin{equation} 2(\Psi_0 +\Psi_0^*) = \frac{1}{PS}  \left[ \left[\frac{P_x}{S}\right]_x +\frac{S_y}{R}\frac{P_y}{R} \right] + \frac{1}{MS} \left[ \left[ \frac{M_x}{S}\right]_x + \frac{S_y}{R}\frac{M_y}{R}  \right] \end{equation}

\begin{equation} 2(-i)(\Psi_0 - \Psi_0^*) = \frac{1}{PR}  \left[ \left[\frac{P_x}{S}\right]_y -\frac{R_x}{S}\frac{P_y}{R} \right] - \frac{1}{MR} \left[ \left[ \frac{M_x}{S}\right]_y -  \frac{R_x}{S}\frac{M_y}{R}  \right] \end{equation}

\begin{equation} 2(-i)(\Psi_0 - \Psi_0^*) = \frac{1}{PS}  \left[ \left[\frac{P_y}{R}\right]_x -\frac{S_y}{R}\frac{P_x}{S} \right] - \frac{1}{MR} \left[ \left[ \frac{M_y}{R}\right]_x -  \frac{S_y}{R}\frac{M_x}{S}  \right] \end{equation}

\begin{equation} 0 = \frac{1}{PR}  \left[ \left[\frac{P_x}{S}\right]_y -\frac{R_x}{S}\frac{P_y}{R} \right] + \frac{1}{MR} \left[ \left[ \frac{M_x}{S}\right]_y -  \frac{R_x}{S}\frac{M_y}{R}  \right] \end{equation}

\begin{equation} 0 = \frac{1}{PS}  \left[ \left[\frac{P_y}{R}\right]_x -\frac{S_y}{R}\frac{P_x}{S} \right] + \frac{1}{MR} \left[ \left[ \frac{M_y}{R}\right]_x -  \frac{S_y}{R}\frac{M_x}{S}  \right] \end{equation}

\begin{equation} \left[\frac{R_x}{S}\right]_x +\left[ \frac{S_y}{R}\right]_y = 0\end{equation}

\subsection{Separation of Hamilton-Jacobi Equation}

It's time to imply the separation of Hamilton-Jacobi equation. As we said before, since our metric functions have no dependency on t,z Hamilton-Jacobi action is soluble with the most simple possible way \cite{carter1968hamilton}. However a more generic separation of variables in Hamilton-Jacobi equation was already achieved by Shapovalov \cite{shapovalov1972separation} and Bagrov \cite{bagrov1991separation} who made known a family of spacetimes with N-parametric Abelian group of motion where N=0,1,2,3.  The HJ action and the corresponding HJ equation are presented.

\begin{equation}\mathcal{S} =at+bz +S_{1}(x) +S_{2}(y)  \end{equation}

\begin{equation}\bar{m}^2 = g^{\mu \nu} \frac{\partial \mathcal{S}}{\partial x^\mu} \frac{\partial \mathcal{S}}{\partial x^\nu} \end{equation}

While the inverse metric is

\begin{equation}g^{\mu \nu} =  \begin{pmatrix}
\frac{P^2 - M^2}{Z^2} & \frac{ML-PN}{Z^2} &0 & 0\\
 \frac{ML-PN}{Z^2} & \frac{N^2-L^2}{Z^2} & 0 & 0\\
0 & 0&   -\frac{1}{S^2} & \\
0 & 0 & 0 & -\frac{1}{R^2}
\end{pmatrix} \end{equation}

Using these previous relations we finally take:

\begin{equation}\bar{m}^2 =  -\frac{\mathcal{S}^2_y}{R^2}-\frac{\mathcal{S}^2_x}{S^2} +\frac{\tilde{B}^2}{M^2}  -\frac{\tilde{A}^2}{P^2}
\end{equation}
The new tilded quantities are constants since they are related with constants of integration A,B and the constants due to the action of the commutative Killing vectors $\partial_t, \partial_z$. 
\begin{equation}\tilde{A} \equiv \frac{a -Ab}{A-B} \end{equation}
\begin{equation}\tilde{B} \equiv \frac{a -Bb}{A-B} \end{equation}

The separation allows us to introduce the function $\Omega^2 \equiv \Phi(x) + \Psi(y)$.

\begin{equation}\Omega^2\bar{m}^2 =  -\frac{\Omega^2}{R^2}\mathcal{S}^2_y-\frac{\Omega^2}{S^2}\mathcal{S}^2_x +\frac{\Omega^2 }{M^2}\tilde{B}^2  -\frac{\Omega^2}{P^2}\tilde{A}^2\end{equation}

Now without loss of generality the separation takes place as 
\begin{equation} \frac{\Omega}{S} = D_S (x) \end{equation}
\begin{equation} \frac{\Omega}{R} = D_R (y) \end{equation}
\begin{equation} \frac{\Omega}{M} = C_M (x) \end{equation}
\begin{equation} \frac{\Omega}{P} = C_P (y) \end{equation}

We shall continue with the resolution of the NPEs (89)-(92) considering the relations (101)-(104).
 
 \begin{equation} \Psi_0 - \Psi_0^* =0 = \left[ \frac{\Omega_x}{\Omega}\right]_y - \frac{\Omega_x}{\Omega}\frac{\Omega_y}{\Omega} \end{equation}
Thus can be rewritten as
 \begin{equation} \Psi_0 - \Psi_0^* =0 =\Phi_x \Psi_y \end{equation}

  In this point we should indicate that there is no essential difference between the two choices that the last relation yields. We choose the option $\Phi_x = 0$. The separation process provides us the relations (101)-(104). Based on the latter and of course on our previous choice we get 
  
  $$R(x,y)\rightarrow R(y) $$
  $$P(x,y) \rightarrow P(y) $$
  
  Thus the real and imaginary parts of Bianchi identity (VI) could be rewritten as follows if we take advantage by the annihilation of the imaginary part of $\Psi_0$. 
 \begin{equation} 2\Psi_0\frac{\Omega_x}{\Omega} - \frac{{C_M}_x}{C_M}[3\Psi_2 + \Psi_0] =0\end{equation}

  \begin{equation} 2\Psi_0\frac{\Omega_y}{\Omega} - \frac{{C_P}_y}{C_P}[3\Psi_2 + \Psi_0] =0\end{equation}
 
 The relation $\Phi(x) = 0 = \Omega_x$  will reform the real part of Bianchi Identity (VI) yielding two possible choices
 \begin{equation}\frac{{C_M}_x}{C_M}[3\Psi_2 + \Psi_0] =0 \end{equation}

In this point we must denote that the annihilation of the bracket is the only acceptable choice.\footnote{Appendix A.}. However our choice along with the equation (108) implies that the $\Omega$ is constant. As an immediate impact we have the following.

\begin{equation}\alpha-\bar{\beta} =0 \end{equation}

 since the $R = R(y)$ and $S=S(x)$ due to the choice that was made during the separation of variables, also the Weyl components are equal to cosmological constant $\Psi_0 = \Psi_4^* = -3\Psi_2 = -3\Lambda$. Thus the only equations that we have to confront are the following.

\begin{equation} 12\Psi_2 = -4\Psi_0 = - \frac{1}{MS} \left[ \frac{M_x}{S} \right]_x \end{equation}

\begin{equation} 12\Psi_2 = -4\Psi_0 = - \frac{1}{PR} \left[ \frac{P_y}{R} \right]_y \end{equation}

One could observe that the two equations are the same if we substitute $M\rightarrow P$ and $S\rightarrow R$. So we may continue with the treatment only of (111). Let's present the non-linear differential equation of second order- we aim to confront- in a most appropriate form.

\begin{equation}\frac{P_{yy}}{P} -\frac{P_y}{P}\frac{R_y}{R} +12\Lambda R^2 = 0 \end{equation}

Without loss of generality we can correlate the two unknown functions with the next relation, where $\Pi$ is a constant of integration.

\begin{equation}\frac{P_{y}}{P} = -\frac{R_y}{R} \rightarrow P(y) = \frac{\Pi}{R(y)} \end{equation}

Thus the equation can be rewritten as follows.

\begin{equation}\frac{R_{yy}}{R} -3 \left( \frac{R_y}{R} \right)^2 -12\Lambda R^2 = 0 \end{equation}

The solution proved to be \footnote{Appendix B.} 

\begin{equation} R^2(y) = \frac{-12\Lambda}{(12\Lambda)^2 (y-C_y)^2 -K} \end{equation}

The latter along with the (114) gives 

\begin{equation} P^2(y) = -\frac{\Pi^2 \left[(12\Lambda)^2 (y-C_y)^2 - K \right]}{12\Lambda} \end{equation}

 In the same fashion we get the corresponding relations for M(x) and S(x).

 \begin{equation} S^2(x) = \frac{-12\Lambda}{(12\Lambda)^2 (x-C_x)^2 - V} \end{equation}

\begin{equation} M^2(x) = -\frac{W^2 \left[(12\Lambda)^2 (x-C_x)^2 - V \right]}{12\Lambda} \end{equation}

Where the quantities $\Pi$,W,K,V,$C_y$,$C_x$ are constants of integration. The constants of integration $C_x, C_y$ have been chosen with a specific manner multiplied by $12\Lambda$, since the annihilation of $\Lambda$ will give us the CF spacetime. We present our metric as follows.
\begin{equation}ds^2 = M^2(x)\left(Adt +dz \right)^2 -P^2(y)\left( Bdt+dz\right)^2 -S^2(x)dx^2 - R^2(y) dy^2 \end{equation}

Applying now the following transformations

\begin{equation} V = 12\Lambda \tilde{V} \end{equation}
\begin{equation} K = 12\Lambda \tilde{K} \end{equation}

 and the substituting the functions of metric we take

\begin{multline}ds^2 =    \left[ \tilde{V} - (12\Lambda) (x-C_x)^2  \right] \left(WAdt +Wdz \right)^2  - \frac{dx^2}{ \tilde{V} - (12\Lambda) (x-C_x)^2 }  \\
 -   \left[\tilde{K} - (12\Lambda) (y-C_y)^2 \right]  \left( \Pi Bdt+\Pi dz\right)^2 -  \frac{dy^2}{ \tilde{K}-(12\Lambda) (y-C_y)^2  }  \end{multline}

\section{Analysis of the Solution}

The metric (123) describes a non-expanding spacetime which is a direct product of 2 two-dimensional spaces of constant curvature. Our metric has already been discovered by Plebanski as his Case C \cite{plebanski1975class}, by Carter as case [$\mathcal{D}$] \cite{carter1968hamilton}, by Hauser and Malhiot as Case (0,0) with two Killing Vectors $\partial_3, \partial_4$ for $\epsilon = +1$ or $\partial_1, \partial_2$ for $\epsilon = -1$ \cite{hauser1978forms}, by Kasner independently \cite{kasner1925algebraic} and by us. This metric is a general family of metrics since it includes Plebanski-Hacyan metric, Bertotti-Robinson and Nariai spacetimes ($\Lambda > 0$)\cite{griffiths2009exact}. 

We shall proceed now with the analysis of the metric. Assuming that the cosmological constant is positive throughout the procedure the constant of integration $\tilde{K}$ must be always positive and the metric must satisfies the following constraint. Otherwise the Lorentzian signature maintenance will be violated.
$$\tilde{K} - (12\Lambda) (y-C_y)^2>0$$

Althought the case for the constant $\tilde{V}$ differs since the $\tilde{V}$ could take both positive and negative values. In case where the constant $\tilde{V}$ is positive the metric (123) holds. The functions $M^2(x)$ and $P^2(y)$ are inversed parabolas, their peak take the values of the constants $\tilde{V},\tilde{K}$ accordingly where the $x =C_x$ and $y=C_y$. Also the squared character of these functions dictates the following constraint. 
$$\tilde{V} - (12\Lambda) (x-C_x)^2>0$$
 The roots of our functions are $x_\pm = C_x\pm \sqrt{\frac{\tilde{V}}{12\Lambda}}$ and $y_\pm = C_y\pm \sqrt{\frac{\tilde{K}}{12\Lambda}}$. The annihilation of the denominators creates coordinate singularities, thus the coordinates x,y lie between the roots i.e inside the positive area of the parabola.
 
 On the other hand where $\tilde{V}$ is negative or zero we confront an interesting case. There is a swap between the character of the coordinates x and t. The x coordinate becomes the timelike coordinate. The metric though has dependence on `time' describing a cosmological model due to the existence of the cosmological constant. In this case x take any value since the functions $M^2(x)$ and $S^2(x)$ are always positive in contrast with the functions of y which take the same values. Then the metric has the following form
 
 \begin{multline}\tag{$123_{CM}$}ds^2 =     \frac{dx^2}{ |\tilde{V}| + (12\Lambda) (x-C_x)^2 } - \left[ |\tilde{V}|+ (12\Lambda) (x-C_x)^2  \right] \left(WAdt +Wdz \right)^2 \\
 -   \left[\tilde{K} - (12\Lambda) (y-C_y)^2 \right]  \left( \Pi Bdt+\Pi dz\right)^2 -  \frac{dy^2}{ \tilde{K}-(12\Lambda) (y-C_y)^2  }  \end{multline}

 \subsection{Reduction to Flat Spacetime}
\textit{Flat Spacetime}

In this section we present two different ways to gain a flat spacetime from metric (123). The annihilation of Weyl Components

\begin{equation} \Psi_0 = \Psi_4^* = - 3\Psi_2  = - 3\Lambda = 0 \end{equation}

 is the standard way to apply this reduction. Indeed, with the appropriate selection of the constants of integration we take

\begin{equation} ds^2 = d\tilde{t}^2 - d\tilde{z}^2 -dx^2-dy^2 \end{equation}

where the constants satisfy the following relations.

\begin{equation}\tilde{V} = 1 =\tilde{K} \end{equation}
\begin{equation}W^2 = 1 \end{equation}
\begin{equation} \Pi^2 = 1\end{equation}

The same reduction is achieved if our constants have the upward values (126)-(128) and in the same time our coordinates $x,y$ takes the values $C_x, C_y$ accordingly (the peak of parabolas $M^2(x),P^2(y)$) since the parenthesis vanishes. 

\subsection{Reduction to Nariai Metric }
The characterization of coordinates in a random metric is not an easy task. There are different paths to determine the nature of each coordinate. Generally a reduction could be proved fruitful in order to determine the generality of solution, additionally reductions are provided in order to gain a deeper understanding about the physical meaning of the resulted metric and the system of coordinates of each metric. Here, we already know by the bibliography that our metric is quite general in a sense since could be able to reduces to various spacetimes. Applying the succeeding transformation 

\begin{equation} \tilde{t} = W(At+z)\end{equation}
\begin{equation} \tilde{z} =\Pi (Bt+z)\end{equation}
\begin{equation} \tilde{x} =  \frac{x-C_x}{\tilde{V}} \end{equation}
\begin{equation} \tilde{y} = \frac{y-C_y}{\tilde{K}}\end{equation}
 
we take a more understandable structure of our spacetime.

\begin{equation}ds^2 =    \left[ 1 -12\Lambda \tilde{x}^2  \right] d\tilde{t}^2  - \frac{d\tilde{x}^2}{ 1 -12\Lambda \tilde{x}^2}  - \left[1-12\Lambda \tilde{y}^2 \right] d\tilde{z}^2 -  \frac{d\tilde{y}^2}{1-12\Lambda \tilde{y}^2 }  \end{equation}
 
The first two-dimensional part of our metric is the two dimensional de-Sitter spacetime of radius $ \frac{1}{\sqrt{12\Lambda}}$.
 
 \begin{equation}ds^2 =    \left[ 1 -12\Lambda \tilde{x}^2  \right] d\tilde{t}^2  - \frac{d\tilde{x}^2}{ 1 -12\Lambda \tilde{x}^2} \end{equation}
 
 In addition if we apply a coordinate transformation in the second part of our metric 

\begin{equation} 12 \Lambda \tilde{z} = \phi\end{equation}
\begin{equation} 12\Lambda \tilde{y} = cos\theta \end{equation}

it can be rewritten as follows.

\begin{equation} ds^2 =\frac{1}{12\Lambda}\left[d\theta^2 +sin^2\theta d\phi^2  \right]   \end{equation}

Thus the latter represents a circle of radius $\frac{1}{\sqrt{12\Lambda}}$. As we know the product $dS_2 \times S^2$ gives the Nariai spacetime. However, it is worth noting that the radius of the metric is related with the cosmological constant which is correlated with the Weyl components and by extension with the tidal forces of the gravitational field. Finally the reduced metric takes the form

\begin{equation}ds^2 =    \left[ 1 -12\Lambda \tilde{x}^2  \right] d\tilde{t}^2  - \frac{d\tilde{x}^2}{ 1 -12\Lambda \tilde{x}^2} -\frac{1}{12\Lambda}\left[d\theta^2 +sin^2\theta d\phi^2  \right] \end{equation}

In this metric it is obvious that as $\tilde{x}$ tends to zero the  first part of metric reduces to two-dimensional flat spacetime. 
A similar analysis can also be found in \cite{griffiths2009exact}.

\section{Geodesics and Constants of Motion}
The equation of geodesics fundamentally describes the phenomenon of the absence of the acceleration that an observer feels along a geodesic line.  Namely, a geodesic line of a gravitational field describes a "free fall" into the gravitational field and can be expressed by the equation of geodesics. In this chapter our focus resides to take advantage of  the symmetries of our problem in order to obtain the Integration Constants of Motion and the geodesic lines in respecto to an affine parameter $\lambda$.

\begin{equation} u^{\mu} u_{\nu;\mu} = 0 \end{equation}

 We define the vector of the observer of mass m as follows

\begin{equation} u^\mu  \equiv \dot{x}^\mu= k_1 n^\mu +k_2 l^\mu+k_3 m^\mu +k_4 \bar{m}^\mu \end{equation}

The derivation of the displacement vector is operated in respect to the affine parameter $\lambda$. The affine parameter is related to the proper time as follows.

\begin{equation} \tau = \bar{m} \lambda \end{equation}

Our Killing tensor is not conformal, hence the only two possible cases which are dictated for the geodesic lines to be are either spacelike or timelike. Additionally the norm of the vector is expressed as follows. The sign (+) is for timelike orbits and the (-) for spacelike orbits.
\begin{equation}k_1 k_2 -k_3 \bar{k}_3 = \pm\frac{1}{2} \end{equation}

Unravelling this we take
\begin{equation}4k_1 k_2 -(k_3 +\bar{k_3})^2 + (k_3 - \bar{k}_3)^2 = \pm4 \end{equation}

The geodesic equation could be easily obtained by the resolution of the Lagrangian equations. The Lagrangian is more suitable for the study of geodesics and is described below.

\begin{equation} \mathcal{L} =\frac{1}{2} g_{\mu \nu} \dot{x}^\mu \dot{x}^\nu \end{equation}

\large 
\textit{Hamilton-Jacobi Action}
\normalsize 

The symmetries of the problem allows us to gain expressions for the 4-vector momentum of the observer implying the separation of variables of the Hamilton-Jacobi equation. Concerning that the coordinates are functions of the affine parameter the action could be expressed as follows.

\begin{equation} \mathcal{S} =\frac{\bar{m}^2}{2}\lambda +at+bz+S_1(x) +S_2(y) \end{equation}

The Hamilton-Jacobi equation is given by 

\begin{equation}\frac{\partial \mathcal{S}}{\partial \lambda} = \frac{1}{2} g^{\mu \nu}\frac{\partial \mathcal{S} }{\partial x^\mu} \frac{\partial \mathcal{S} }{\partial x^\nu}\end{equation} 

In case where the function $\Omega$ is constant the relation (100) could be rewritten downward.

\begin{equation}\bar{m}^2 =   -\frac{\mathcal{S}^2_y}{R^2(y)}-\frac{\mathcal{S}^2_x}{S^2(x)} +\frac{\tilde{B}^2}{M^2(x)}  -\frac{\tilde{A}^2}{P^2(y)}\end{equation}

\begin{equation}\tilde{A} \equiv \frac{a -Ab}{A-B} \end{equation}
\begin{equation}\tilde{B} \equiv \frac{a -Bb}{A-B} \end{equation}

The definition of the constant $\mathcal{K}$ is dictated by the separation of variables in Hamilton-Jacobi equation.  

\begin{equation} \mathcal{K} \equiv \frac{\mathcal{S}^2_y}{R^2(y)} + \frac{\tilde{A}^2}{P^2(y)} + \frac{\bar{m}^2}{2} = -\frac{\mathcal{S}^2_x}{S^2(x)} + \frac{\tilde{B}^2}{M^2(x)} - \frac{\bar{m}^2}{2} \end{equation}

The canonical momentum is correlated with the vector of the observer as follows.

\begin{equation} p_\mu = g_{\mu \nu} u^\nu \end{equation}

Next imposing the relations of the metric components along with the definition of the velocity of the observer in the previous relations  

\begin{equation}p_t = \left[A^2 M^2(x) - B^2 P^2(y) \right]\dot{t} + \left[ AM^2(x)-BP^2(y)\right] \dot{z} \end{equation}
\begin{equation}p_z = \left[ AM^2(x)-BP^2(y)\right] \dot{t} +\left[ M^2(x) - P^2(y) \right]\dot{z} \end{equation}
\begin{equation}p_x = -S^2(x) \dot{x} \end{equation}
\begin{equation}p_y = -R^2(y) \dot{y} \end{equation}

The normalizing condition of the system is equivalent with the conservation of the rest mass.

\begin{equation}\bar{m}^2 =  g_{\mu \nu} u^\mu u^\nu \end{equation}

Along these lines the Hamiltonian is defined by

\begin{equation} \mathcal{H} = \frac{1}{2} g_{\mu \nu} u^\mu u^\nu\end{equation}

 The Hamiltonian is a conserved quantity of the problem since is correlated with the conserved rest mass. Furthermore the momentum is the derivative of the action. Hence using the relation (150) we take expressions for $p_x,p_y$ \footnote{The sign of the square roots could be chosen independently although for reasons of convenience we take the positive sign for both cases.}. Considering that the components of the 4-vector momentum is the partial derivative of the action with respect to coordinates then it could be expressed as follows.

\begin{equation}p_\mu = \left( a,b, S(x) \left[\frac{\tilde{B}^2}{W^2} S^2(x)- \mathcal{K} +\frac{\bar{m}^2}{2} \right]^{1/2} , R(y)\left[ \mathcal{K} +\frac{\bar{m}^2}{2} -\frac{\tilde{A}^2}{\Pi^2}R^2(y)\right]^{1/2}  \right)\end{equation}

The comparison between the latter and the relations (152)-(155) resulting to the geodesic equations
\begin{equation}a = \left[A^2 M^2(x) - B^2 P^2(y) \right]\dot{t} + \left[ AM^2(x)-BP^2(y)\right] \dot{z} \end{equation}
\begin{equation}b = \left[ AM^2(x)-BP^2(y)\right] \dot{t} +\left[ M^2(x) - P^2(y) \right]\dot{z} \end{equation}
\begin{equation}  S(x) \left[\frac{\tilde{B}^2}{W^2} S^2(x)- \mathcal{K} -\frac{\bar{m}^2}{2} \right]^{1/2}= -S^2(x) \dot{x} \end{equation}
\begin{equation}R(y)\left[ \mathcal{K} -\frac{\bar{m}^2}{2} -\frac{\tilde{A}^2}{\Pi^2}R^2(y)\right]^{1/2} = -R^2(y) \dot{y} \end{equation}

Trying to find expressions for $\dot{t}, \dot{z}$ we multiply the relation (160) with B and A accordingly and make a subtraction with (159).

\begin{equation} A \dot{t} +\dot{z}= \frac{\tilde{B}}{M^2(x)} = \frac{\tilde{B}}{W^2\left[\tilde{V} -12\Lambda(x-C_x)^2\right]} \end{equation} 

\begin{equation} B \dot{t} +\dot{z}= \frac{\tilde{A}}{P^2(y)} =\frac{\tilde{A}}{\Pi^2\left[\tilde{K} -12\Lambda(y-C_y)^2\right]}   \end{equation}

These relations makes clear that the succeeding transformation would be proved useful. 

\begin{equation}\tilde{t} = W^2(At+z) \end{equation}
\begin{equation}\tilde{z} = \Pi^2(Bt+z) \end{equation}

Thus the relation (163) and (164) are

\begin{equation}  \dot{\tilde{t}} = \frac{\tilde{B}}{\tilde{V} -12\Lambda(x-C_x)^2} \end{equation} 

\begin{equation}  \dot{\tilde{z}} = \frac{\tilde{A}}{\tilde{K} -12\Lambda(y-C_y)^2} \end{equation}

Considering the expressions for $S^2(x)$ and $R^2(y)$ the relations (161) and (162) can be treated as follows. It has been convenient to define a new constant $\mathcal{K}_{\pm} \equiv \mathcal{K}\pm\frac{\bar{m}^2}{2}$

\begin{equation} -\frac{dx}{d\lambda} = \left[ \frac{\tilde{B}^2}{W^2} - \mathcal{K}_+  \left[ \tilde{V} - 12\Lambda(x-C_x)^2 \right] \right]^{\frac{1}{2}} \end{equation}

\begin{equation} \rightarrow \frac{dx}{\left[  1- \frac{12\Lambda W^2 \mathcal{K}_+}{\tilde{V} W^2 \mathcal{K}_+ -\tilde{B}^2} (x-C_x)^2 \right]^{\frac{1}{2}}} =  \left[\frac{  \tilde{V}W^2\mathcal{K}_+ -\tilde{B}^2}{W^2}\right]^{\frac{1}{2}}  d\lambda \end{equation}

Next we need to multiply with $\sqrt{\frac{\mathcal{K}_+ 12\Lambda W^2}{\mathcal{K}_+ \tilde{V}W^2 - \tilde{B}^2}}$ and integrate both parts.

\begin{equation}  arcsin\left( \sqrt{\frac{\mathcal{K}_+ 12\Lambda W^2}{\mathcal{K}_+ \tilde{V}W^2 - \tilde{B}^2}}(x-C_x) \right) =\sqrt{\mathcal{K}_+ 12\Lambda} \lambda+s_0  \end{equation}

Finally the $x,y$ in respect to $\lambda$ is presented.

\begin{equation} x(\lambda) = \sqrt{\frac{\mathcal{K}_+ \tilde{V}W^2 - \tilde{B}^2}{\mathcal{K}_+ 12\Lambda W^2}}sin\left( \sqrt{\mathcal{K}_+ 12\Lambda} \lambda +s_0 \right) +C_x \end{equation}

\begin{equation} y(\lambda) = \sqrt{\frac{\mathcal{K}_- \tilde{K} \Pi^2 - \tilde{A}^2}{ \mathcal{K}_- \Pi^2 12\Lambda}} sin\left( r_0 - \sqrt{\mathcal{K}_- 12\Lambda}\lambda \right) +C_y \end{equation}

We have already expressed the coordinates x,y in respect to the affine parameter $\lambda$ where the $s_0, r_0$ are constants of integration. We shall continue implying the last expressions into (169) and (170) accordingly. At last we present our new coordinates in respect to affine parameter.

\begin{equation} \tilde{t} (\lambda) = - \frac{W^2}{\tilde{B}}\sqrt{\frac{\mathcal{K}_+}{12\Lambda}}  \left[\frac{cos\left( \sqrt{\mathcal{K}_+ 12\Lambda} \lambda +s_0\right)}{sin\left( \sqrt{\mathcal{K}_+ 12\Lambda} \lambda +s_0\right)} -C_0 \right] \end{equation}

\begin{equation} \tilde{z} (\lambda) =  \frac{\Pi^2}{\tilde{A}}\sqrt{\frac{\mathcal{K}_-}{12\Lambda}}  \left[\frac{cos\left( r_0 -\sqrt{\mathcal{K}_- 12\Lambda} \lambda \right)}{sin\left( r_0 - \sqrt{\mathcal{K}_- 12\Lambda} \lambda  \right)} -\tilde{C_0} \right] \end{equation}

We observe that during the geodesic line when the affine parameter reaches the values

\begin{equation}\lambda =  \frac{- s_0}{\sqrt{\mathcal{K}_+ 12\Lambda}} = \frac{r_0}{\sqrt{\mathcal{K}_- 12\Lambda}}  \end{equation}

then the x$\rightarrow C_x$ and y$\rightarrow C_y$ and of course the relations (174) and (175) determines that the ignorable tilded coordinates goes to infinity.

Our geodesics of the non-ignorable coordinates are complete since the affine parameter could reach any real value due to the bounded character of the relations (172)-(175). Furthermore the relations (172) and (173) clarify that the geodesics of non-ignorable shifted coordinates are bounded. However when the affine parameter approaches the values of the relation (176) the equations of motion of the ignorable coordinates goes to infinity. The sign of the infinity depends from the sign of the sin() function. Of course we have already made a choice about the sign during the integration of geodesics. The latter must be considered in order to determine the sign of $\tilde{t}, \tilde{z}$. As we stated in subsection (6.1) when the affine parameter reaches these points our coordinates x,y take the values $C_x,C_y$ in the same time.  
Our spacetime with the above transformations coincide with the bespoken spacetimes. Hence the trajectories of the observer lies on the merging between the hyperboloid and the $S^2$ sphere with the appropriate choice of the constants of integration. Besides both characterized by the same curvature which is equal to $\frac{1}{\sqrt{12\Lambda}}$.

At last it should be denoted that the transformation (165),(166) could help us to revert in the previous non-tilded coordinates but the analysis as well as the vulnerable points are the same.

\subsection{Killing Tensor and Constants of Motion}

Since the beginning of the analysis there was any connection connecting the role of the eigenvalues of the Killing Tensor with the constants of motion of the problem.

The real parts of the reformed relations (13) and (16) have the following form.

\begin{equation} (\delta+\bar{\delta}) \lambda_0 = 2\left[ \lambda_0(\pi+\bar{\pi}) - (\kappa+\bar{\kappa})(\lambda_1+\lambda_2) \right] \end{equation}  
\begin{equation} (\delta+\bar{\delta}) \lambda_0 = 2\left[ \lambda_0(\bar{\pi}-\pi) - (\kappa-\bar{\kappa})(\lambda_1+\lambda_2) \right] \end{equation}
  \begin{equation} (\delta+\bar{\delta}) \lambda_1 = -2\left[ \lambda_0(\kappa+\bar{\kappa}) - (\pi+\bar{\pi})(\lambda_1+\lambda_2) \right] \end{equation} 
   \begin{equation} (\delta+\bar{\delta}) \lambda_1 =- 2\left[ \lambda_0( \kappa-\bar{\kappa}) - (\bar{\pi} - \pi)(\lambda_1+\lambda_2) \right] \end{equation}

The above relations resulted to the downward relations with $\lambda_\pm$ are constants of integration. The non-constant eigenvalues of the Killing Tensor are the following.

\begin{equation} \lambda_0 +\lambda_1 = \lambda_+ M^2(x) = \lambda_+ W^2 \left[ \tilde{V} - 12\Lambda (x-C_x)^2 \right]  \end{equation}
\begin{equation} \lambda_0 -\lambda_1 = \lambda_- P^2(y) = \lambda_-  \Pi^2 \left[ \tilde{K} -12\Lambda (y - C_y)^2 \right]  \end{equation}

 It is clear now that our eigenvalues depend by the functions $M^2(x),P^2(y)$ and by extension by the non-ignorable coordinates. Actually the analysis of the eigenvalues of the Killing Tensor interest us in three cases. The first two cases concern the vulnerable points of the functions of metric. These points are $x\rightarrow C_x$ and $y\rightarrow C_y$ and the poles  $x_\pm = C_x\pm \sqrt{\frac{\tilde{V}}{12\Lambda}}$ and $y_\pm = C_y\pm \sqrt{\frac{\tilde{K}}{12\Lambda}}$. The corresponding values of the  first case are constant, in this case the inversed parabola reaches the peak hence $\lambda_0 +\lambda_1 = \lambda_+  \tilde{V}$ and$\lambda_0 -\lambda_1 = \lambda_- \tilde{K}$. On the other hand when the coordinates reach the poles the eigenvalues tend to zero.
 
   Next we shall determine the Fourth Constant of motion using the following relation.

\begin{equation} K^{\mu \nu} p_\mu p_\nu = \mathcal{K}   \end{equation}

The inverse Killing Tensor is 

\begin{equation}K^{\mu\nu} = \begin{pmatrix}
    \frac{\lambda_0}{\lambda^2_0 -\lambda^2_1}  & -\frac{\lambda_1}{\lambda^2_0 -\lambda^2_1} & 0 & 0 \\
    -\frac{\lambda_1}{\lambda^2_0 -\lambda^2_1}   & \frac{\lambda_0}{\lambda^2_0 -\lambda^2_1} & 0 & 0 \\
    0 & 0 &  0 & -\frac{1}{\lambda_2} \\
    0 & 0 & -\frac{1}{\lambda_2} & 0 \\
    \end{pmatrix} 
\end{equation}

while the vector of the observer is given by the relation (158).
\begin{equation}p_\mu =  \left( a,b, S(x) \left[\frac{\tilde{B}^2}{W^2} S^2(x)- \mathcal{K}_+ \right]^{1/2} , R(y)\left[ \mathcal{K}_- -\frac{\tilde{A}^2}{\Pi^2}R^2(y)\right]^{1/2}  \right) \end{equation}

 The relation (184) along with the (185) and (186) resulting to
 
 \begin{equation} \frac{1}{2} \left[  \frac{(a-b)^2}{\lambda_0-\lambda_1}+\frac{(a+b)^2}{\lambda_0 +\lambda_1} \right] -\frac{R(y) S(x)}{\lambda_2}  \sqrt{\left(\frac{\tilde{B}^2}{W^2} S^2(x)- \mathcal{K}_+ \right) \left( \frac{\tilde{A}^2}{\Pi^2}R^2(y) -\mathcal{K}_- \right)}= \mathcal{K} \end{equation}

Additionally the third case that we didin't referred to concerns two exceptional points. These points annihilate the square root part of the relation (186). 
 
 \begin{equation}x_0=C_x \pm \sqrt{\frac{\tilde{V} \mathcal{K}_+ W^2 - \tilde{B}^2}{12\Lambda \mathcal{K}_+W^2}}\end{equation}
 \begin{equation}y_0=C_y \pm \sqrt{\frac{\tilde{K} \mathcal{K}_- \Pi^2 - \tilde{A}^2}{12\Lambda \mathcal{K}_- \Pi^2}}\end{equation}

 These points give relations between the eigenvalues and the constants of motion of course with the use of (186). Using the (187) first in (181) and both to (186) we take
 \begin{equation}\tag{181}\lambda_0 + \lambda_1 = \lambda_+ \frac{\tilde{B}^2}{\mathcal{K}_+ W^2 }  \end{equation}

\begin{equation}\lambda_0 - \lambda_1 = \frac{(a-b)^2 \lambda_+ \tilde{B}^2}{2\mathcal{K} \lambda_+\tilde{B}^2 - (a+b)^2 \mathcal{K}_+ W^2} \end{equation}
 
same as before, when the coordinate $y\rightarrow y_0$ then the eigenvalues take the following form.

\begin{equation}\tag{182}\lambda_0 - \lambda_1 = \lambda_- \frac{\tilde{A}^2}{\mathcal{K}_- \Pi^2 }  \end{equation}

\begin{equation}\lambda_0 + \lambda_1 = \frac{(a+b)^2 \lambda_- \tilde{A}^2}{2\mathcal{K} \lambda_-\tilde{A}^2 - (a-b)^2 \mathcal{K}_- \Pi^2} \end{equation}

\section{Discussion and Conclusions}

This work is the initial part of a study endeavors to establish a comprehensive understanding about the spacetimes in vacuum which admit the canonical forms of the Killing Tensor. The involution of the Canonical Forms in the resolution process of the Einstein's Field Equations proved fruitful providing a vast variety of solutions at first glance of all geometrical types except Type II. These types of solutions admit the 2nd and 3rd Canonical Form of the Killing Tensor. Essentially the rotation we applied scoping to simplify the problem giving birth to the three classes that contain all these types of solutions. Specifically the simplification via the rotation could be operated due to the initial annihilation of the eigenvalue $\lambda_7$, in case where the terms $\tilde{\theta}^3 \otimes \tilde{\theta}^3$ and $\tilde{\theta}^4 \otimes \tilde{\theta}^4$ were present along with the invariant character of the Killing tensor would annihilate the remaining free parameter of the rotation, the parameter $b$. Strictly speaking the capitalization of the parameter $b$ manage to gain the $\textbf{Key relations}$ which basically determine our solutions. Thus a rotation around a null tetrad frame could provide multiple solutions only if a reduced form of the 2nd and 3rd full Canonical form of the Killing Tensor is considered. 

The novelty in this work is that the Carter's Case [$\mathcal{D}$] and the cosmological model that derived admit the 2nd Canonical form of Killing Tensor which contains two non-constant eigenvalues that depends on quadratic functions of the non-ignorable coordinates and one eigenvalue proved to be constant. We are already aware that Carter's Family of Metrics admits a Killing Tensor \cite{hauser1978forms}. 

It has already been stated that there are various Killing Tensors that could be used as initial premise of symmetry in order to obtain exact solutions from EFEs. Aiming to find general families of solutions, we initiated by the fact that the Canonical form of the Killing Tensor are the most general form of a Killing Tensor since involves four distinct eigenvalues, thus our Killing Tensor is a generalization of a diagonalized one with two double eigenvalues. Although from Hauser-Malhiot's work\cite{hauser1978forms} is proven that the diagonalized Killing Tensor with two double eigenvalues admitted by the Carter's Case [$\mathcal{D}$] as case (0,0) with $\epsilon = +1$, without the multiplied integration factor and with two Killing vectors $\partial_3, \partial_4$.

Along these lines there is a degeneracy of the Killing Tensor admitted by the Carter's Case [$\mathcal{D}$] and the question that comes to surface is how the admission of the existence of a more general Killing Tensor could not provide us with more general families of solution? One could answer that the overdetermination of the system via the integrability conditions and the implication of the rotation sometimes would provide us with quite special solutions. We do not have a formal answer in this moment. Althought we believe that the concept of degeneracy of Killing Tensor answers partly. 

On the other hand the fact that the corresponding eigenvalues $\tilde{\lambda}_1, \tilde{\lambda}_2$ of the diagonalized Killing tensor of the bespoken article could have a similar quadratic form of the non-ignorable coordinates with our eigenvalues $\lambda_0 \pm \lambda_1$ giving birth to the following question. Does exist any correlation between these two Killing Tensors and by extension, does exist any combination of rotations around the null tetrad frames in which the two Killing Tensors coincide? 
 
$${\tilde{K}{^{HM}}}_{\mu \nu} =  \mathcal{R} {K{^2}}_{\mu \nu} $$

$$ \rightarrow \begin{pmatrix}
0 & \tilde{\lambda_1}  & 0 & 0\\
 \tilde{\lambda_1} & 0  & 0 & 0\\
0 & 0&  0   & \tilde{\lambda_2} \\
0 & 0 & \tilde{\lambda_2} & 0   
\end{pmatrix} = \mathcal{R}  \begin{pmatrix}
0& \lambda_0 + \lambda_1  +C_1 \lambda_2   & 0 & 0\\
\lambda_0 + \lambda_1  +C_1 \lambda_2 & 0 & 0 & 0\\
0 & 0& 0   &\lambda_0 - \lambda_1  +C_2 \lambda_2 \\
0 & 0 & \lambda_0 - \lambda_1  +C_2 \lambda_2 & 0
\end{pmatrix}$$

The answer though is negative. A first rotation around the $l_\mu$ tetrad frame scatters away the elements of $K^2_{\mu \nu}$ in every element of the tensor multiplied with the components of the rotation. A second rotation around $n_\mu$ tetrad will do exactly the same but this time we also have in account the components of the second rotation. Regarding the two rotations we hoped that the appropriate arrangement of the parameters of the rotations would be the key to correlate these two tensors but this proved impossible. The lengthy proof is not presented here. 

At last we ought to comment that considering the collection of our spin coefficients we conclude that the Gordberg-Sachs theorem is not true in our case. Indeed, Debever in \cite{debever1984exhaustive} states that when the condition $\Psi_0 \Psi_4 =9\Psi_2^2$ satisfied then the Generalized Goldberg-Sachs Theorem does not hold. However there is an exceptional case has been studied by Plebanski and Hacyan \cite{Plebanski1979some} and Garcia and Plebanski \cite{Garcia1982multiexponent} where there are metrics like ours.

\section{Appendix A}
In this Appendix we going to analyze the outcomes of the other cases that the following equation yields. 

$$ \frac{{C_M}_x}{C_M}(3\Psi_2+\Psi_0)=0 $$

The latter yields two cases.

Case I: ${C_M}_x = 0 \neq 3\Psi_2+\Psi_0 $

Case II: ${C_M}_x = 0 = 3\Psi_2+\Psi_0 $

Let us remind to the reader that we already are aware that $P,R$ depends only on y since with $\Psi_0 =\Psi_0^*$ we take the annihilation of $\Phi(x)$. Now the other choice of the relation (111) implies that $M(x,y) \rightarrow M(y)$. Hence the contribution that one could gain from NPEs (70)-(78) and BI (VI) is the following.
\begin{equation}\tag{A.1} 12\Psi_2 = -\frac{1}{PR}  \left[\frac{P_y}{R}\right]_y  - \frac{1}{MR} \left[ \frac{M_y}{R}\right]_y  \end{equation}

\begin{equation}\tag{A.2} 12\Psi_2 = -\frac{S_y}{RS}  \left[  \frac{P_y}{PR} +\frac{M_y}{MR} \right]  \end{equation}

\begin{equation}\tag{A.3} 4\Psi_0 = \frac{1}{PR}   \left[\frac{P_y}{R}\right]_y - \frac{1}{MR}  \left[ \frac{M_y}{R}\right]_y  \end{equation}

\begin{equation}\tag{A.4} 4\Psi_0 = \frac{S_y}{RS}  \left[\frac{P_y}{PR} +\frac{M_y}{MR}   \right]  \end{equation}

\begin{equation}\tag{A.5} \left[ \frac{S_y}{R}\right]_y=0 \end{equation}

\begin{equation}\tag{A.6}  {C_M}_x =0 \end{equation}

\begin{equation}\tag{A.7} 2\Psi_0\frac{\Omega_y}{\Omega} - \frac{{C_P}_y}{C_P}[3\Psi_2 + \Psi_0] =0\end{equation}
 
If we add (A.1) with (A.3) and (A.2) with (A.4) accordingly we take

\begin{equation}\tag{A.8} 3\Psi_2 + \Psi_0  =0 = \left[ \frac{M_y}{R}\right]_y \end{equation} 

The last expression clarifies that the Weyl component $\Psi_0$ is also constant so the Case I is impossible. Hence we continue the analysis only for Case II. 

The imaginary part of BI (VI) which is expressed by relation (182) along with the latest annihilation dictates that $\Omega_y = 0$ which makes the metric function M(y) constant. In addition the metric function $S(x,y) \rightarrow S(x)$ since the only contribution in respect to y vanished along with $\Omega_y$.  According to the latter the relations (177) and (179) makes our spacetime conformally flat resulting to 
\begin{equation}\tag{A.9}\Psi_2 = \Psi_0 = \Psi_4 = 0 \end{equation}

\vspace{5cm}
\section{Appendix B}
Thus the only equations that we have to confront are the following.
\begin{equation} 12\Psi_2 = -4\Psi_0 = - \frac{1}{MS} \left[ \frac{M_x}{S} \right]_x \end{equation}

\begin{equation} 12\Psi_2 = -4\Psi_0 = - \frac{1}{PR} \left[ \frac{P_y}{R} \right]_y \end{equation}

One could observe that the two equations are the same if we substitute $M\rightarrow P$ and $S\rightarrow R$. So we may continue with the treatment only of (113). Let's present the non-linear differential equation of second order in a most usable form.

\begin{equation}\frac{P_{yy}}{P} -\frac{P_y}{P}\frac{R_y}{R} +12\Lambda R^2 = 0 \end{equation}

Without loss of generality we can correlate the two unknown functions with the next relation, where $\Pi$ is a constant of integration.

\begin{equation}\frac{P_{y}}{P} = -\frac{R_y}{R} \rightarrow P(y) = \frac{\Pi}{R(y)} \end{equation}

Thus our equation is a non-linear differential equation of second order 
$$  \frac{R_{yy}}{R} -3\left( \frac{R_y}{R} \right)^2 -12\Lambda R^2 = 0  $$

in order to begin its resolution we have to define the function.

$$k\equiv \frac{dR(y)}{dy} $$

Then the derivative of k with respect to R could be obtained by the first derivative with respect to y

$$\frac{dk}{dy} = \frac{dk}{dR}\frac{dR}{dy} \rightarrow k_R k =  R_{yy}   $$
 then the differential equation could be rewritten as follows
 
 $$ (k^2)_{R} -6\frac{k^2}{R} -24 \Lambda R^3 = 0  $$
 
 Next, we can divide our function to a homogeneous solution and to a partial solution. In this case these indices do not indicate derivation.
 $$k^2 = k_0^2 + k_P^2$$ 
 
\textbf{Homogeneous Solution}: ${(k_0^2)}_R - 6 \frac{(k_0)^2}{R}=0 \rightarrow k_0^2 = K R^6$ where K is constant.

\textbf{Partial Solution}: ${(k_P^2)} = \tilde{U} R^4$ where $\tilde{U}$ is also a constant.

If we substitute our solution in the differential equation we take 

$$k^2 = KR^6 - 12\Lambda R^4 \rightarrow k = -e R^2 \sqrt{KR^2 - 12\Lambda} $$

In this poing we define $e\equiv \pm$. In order to express R as a function of y we have to proceed backwards considering that $k\equiv \frac{dR(y)}{dy}$. Afterwards we take the following integral.

$$\frac{dR}{R^2 \sqrt{KR^2-12\Lambda}} =  -e dy$$

Applying the following transformation to the left part of the integral 

$$\sqrt{\frac{K}{12\Lambda}} R \equiv cosw$$

then we take

$$ \sqrt{\frac{K}{12\Lambda}} \frac{dw}{cos^2w} = e\sqrt{12 \Lambda} dy   $$

After the integration the result is
 
 $$ \sqrt{\frac{K}{12\Lambda}} tanw = e\sqrt{12 \Lambda}(y-C_y)   $$

and with the usage of $\sqrt{\frac{K}{12\Lambda}} R = cosw$ we finaly take
 
$$ R^2(y) = \frac{-12\Lambda}{(12\Lambda)^2 (y-eC_y)^2 - K}$$

As one could observe e doesn't have any special contribution since its existence equivalently means just a shift on the x axis. Hence we consider it to be equal +1. At last, we obtained the corresponding solution for $M^2(x)$ with exactly the same manner since the initial differential equations are the same. The integration constant is multiplied by $12\Lambda$. With this choice the annihilation of the cosmological constant reduce our spacetime in Minkowski flat spacetime with the appropriate choice of the constants.



\begin{thebibliography}{0}    



\bibitem{cariglia2014hidden}M. Cariglia, {\it Reviews of Modern Physics}, {\bf 86}, (2014) 1283.
 
\bibitem{taxiarchis1985space}T. Papakostas, {\it General Relativity and gravitation}, {\bf 17}, (1985) 149.
 
 \bibitem{hauser1976space}I. Hauser and RJ. Malhiot, {\it J. Math. Phys.}, {\bf17}, (1976) 1306.

\bibitem{carter1968hamilton}B. Carter, {\it Commun. Math. Phys.}, {\bf 10}, (1968) 280
 
\bibitem{petrov2000classification}AZ Petrov, {\it General Relativity and Gravitation}, {\bf32}, (2000) 1665.


\bibitem{newman1962approach}E Newman and R Penrose, {\it J. Math. Phys.}, {\bf 3}, (1962) 566.

\bibitem{debeverriemann}R. Debever,  {\it The Riemann tensor in general relativity}, University of Brussels (1964). 
 
\bibitem{cahen1967complex}M. Cahen, R. Debever and L. Defrise {\it J. Math. and Mech.}, {\bf16}, (1967) 761.

\bibitem{griffiths2009exact}J. Griffiths and J. Podolsky, {\it Exact Space-Times in Einstein's General Relativity} (Cambridge University Press, 2009) 

\bibitem{stephani2009exact} Stephani, Hans and Kramer, Dietrich and MacCallum, Malcolm and Hoenselaers, Cornelius and Herlt, Eduard,  {\it Exact solutions of Einstein's field equations}, (Cambridge university press,2009)
 
 \bibitem{chandrasekhar1986new} S. Chandrasekhar and BC Xanthopoulos, {\it Proc. Royal Soc. London A. Math. Phys. Sc.}, {\bf408}, (1986) 175.
 
 \bibitem{debever1981orthogonal}R. Debever and RG McLenaghan, {\it J. Math. Phys.}, {\bf22}, (1981) 1711.
 
 \bibitem{shapovalov1972separation}VN Shapovalov, VG Bagrov, AG Meshkov, {\it Soviet Physics Journal}, {\bf 15}, (1972) 1115.
 
 \bibitem{bagrov1991separation}VG Bagrov, AV Shapovalov and AA Yevseyevich, {\it Classical and Quantum Gravity}, {\bf 8}, (1991) 163.
 
  \bibitem{debever1984exhaustive}R. Debever, N. Kamran and RG McLenaghan, {\it Journal of mathematical physics}, {\bf25}, (1984) 1955.
 
 \bibitem{plebanski1975class}JF Plebanski, {\it Annals of Physics}, {\bf 90}, (1975) 196.
 
\bibitem{hauser1978forms}I. Hauser and RJ. Malhiot, {\it J. Math. Phys.}, {\bf 19}, (1978) 187.

\bibitem{kasner1925algebraic}E. Kasner, {\it Transactions of the Amer. Math. Soc.}, {\bf 27}, (1925) 101.
  
 \bibitem{debever1979riemannian}R. Debever, RG McLenaghan and N. Tariq, {\it General Relativity and Gravitation} {\bf10} (1979) 853.

 \bibitem{Plebanski1979some}J.F. Plebanski and S. Hacyan, {\it J. Math. Phys.} {\bf20}, (1979) 1004.
 \bibitem{Garcia1982multiexponent}A. Garcia and J.F. Plebanski, {\it J. Math. Phys.} {\bf23}, (1982) 123.


 




\end{thebibliography}
\end{document}